\documentclass[pre]{revtex4}
\usepackage{graphicx}
\usepackage{graphics}
\begin{document}

\noindent
\leftline{\scriptsize{From: \textbf{\textit{Quantum Annealing and Related Optimization Methods}}}}\\
\leftline{\scriptsize{A. Das and B. K. Chakrabarti (Eds.), LNP {\bf{679}}, pp 3-36, Springer, 
Hidelberg (2005).}}

\medskip

\title{Transverse Ising Model, Glass and Quantum Annealing}

\author{Bikas K. Chakrabarti$^1$ and Arnab Das$^2$}
\affiliation{Theoretical Condensed Matter Physics Division and Center for
Applied Mathematics and Computational Sciences, Saha Institute of Nuclear 
Physics, 1/AF, Bidhannagar, Kolkata, India \\
\texttt{$^1$e-mail: bikask.chakrabarti@saha.ac.in}\\
 \texttt{$^2$e-mail: arnab.das@saha.ac.in}}
%
%
\maketitle

%
%
%

%

\section{Introduction}

 In many physical systems, cooperative interactions between 
spin-like (two-state) degrees of freedom tend to establish some kind of order
in the system, while the presence of some noise effect (due to
temperature, external transverse field etc.) tends to destroy it. 
Tranverse Ising model can  quite
succeessfully be employed to study the order-disorder transitions in
many of such systems.

An example of the above is the study of ferro-electric
 ordering in Pottasium Dihydrogen Phosphate (KDP) type systems
 (see, e.g., \cite{BC:Blinc}). 
To understand such ordering, the basic structure
can be viewed as a lattice, where in each lattice  point there is a 
double-well potential created by an
 oxyzen atom and the hydrogen or proton resides within it 
in any of the two wells.
 In the corrosponding
 Ising (or pseudo-spin) picture the state of a double-well with a
 proton at the left-well and that with one at the right-well are represented
 by, say, $|\uparrow\rangle$ and $|\downarrow\rangle$ respectively 
(see, for a portion of the lattice, Fig. 1.1).
\begin{figure}
\centerline{\includegraphics[width=0.6\textwidth, angle=270]{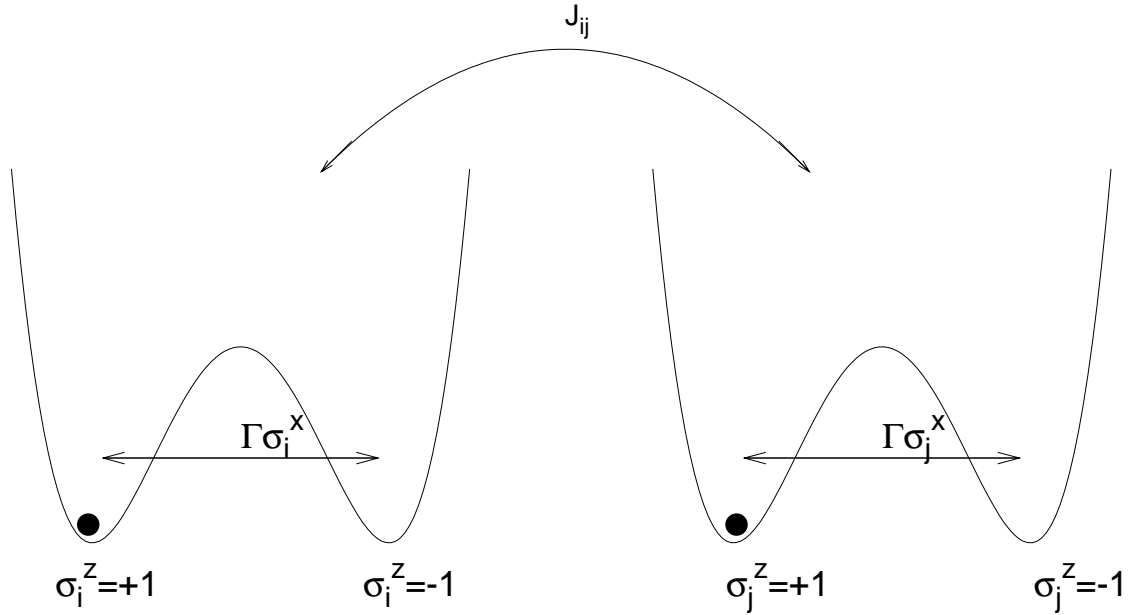}}
\caption{\small{The double wells at each site (e.g., provided by 
oxygen in KDP) provide two (low-lying) states of the proton (shown by each
 double well) indicated by the Ising states $|\uparrow\rangle$ and 
$|\downarrow\rangle$ at each site. The tunnelling between the states are
induced by the transverse field term ($\Gamma\sigma^{x}$).
 The dipole-dipole interaction 
$J_{ij}$ here for the (asymmetric) choice of one or the other well at each
site induces the `exchange' interaction as shown.}}
\end{figure}
\noindent The protons at neighbouring sites have mutual dipolar
 repulsions.
Hence had proton been a classical particle, the zero -temperature 
configuration of the
 system would be one with either
 all the protons residing  at their respective left-well or
 all residing at the right-well (corrosponding to the all-up or all-down
 configuration of the spin system in presence of cooperative interaction alone,
at zero-temperature). Considering no fluctuation at zero temperature, 
the Hamiltonian for the system in the corrosponding pseudo-spin picture will  
 just be identical to the classical Ising 
Hamiltonian (without any transverse term). 
However, proton being a quantum particle, there is always a 
finite probability for it to tunnel through the finite barrier between
 two wells even at zero-temperature due to quantum fluctuations.
To formulate the term for the tunnelling in the corrosponding spin-picture, we 
notice that $\sigma^x$ is the right operater. This is because
 $$\sigma^x |\uparrow\rangle
 = |\downarrow\rangle\quad \mathrm{and}\quad \sigma^x|\downarrow\rangle = 
|\uparrow\rangle, \eqno (1) $$ 

\noindent where $|\uparrow\rangle$ represents the state where
 the proton is in the left 
well, while $|\downarrow\rangle$ represents that with the proton in the 
right well.
           Hence the tunelling term will  exactly be represented by the
 tranvere field term in the transverse Ising Hamiltonian. Here the
 transverse field coefficient $\Gamma$ will represent the tunnelling 
integral, which depends on the width and height of the barrier,
 mass of the particle, etc.\\

\section{Transverse Ising Model (TIM)} 

Such a system as discussed above, can be represented by a quantum
 Ising system, having Hamiltonian
$$\mathcal{H} = -\sum_{\langle i,j\rangle } J_{ij} \sigma_{i}^z 
\sigma_{j}^z -\Gamma\sum_{i}\sigma_{i}^{x}. \eqno (2)$$

\noindent Here, $J_{ij}$ is the coupling between the spins at 
sites $i$ and $j$, where
 $\sigma^{\alpha}$'s ($\alpha = x,y,z$) are the Pauli spins satisfying the
 commutation relations 
$$ [\sigma_{i}^{\alpha} , \sigma_{j}^{\beta} ] = 2i\delta_{ij} \epsilon_{\alpha
\beta\gamma} \sigma_{i}^{\gamma} \eqno (3) $$
\noindent Here, $\delta_{ij}$ is the
 Kr\"{o}necker's $\delta$, and $\epsilon_{\alpha\beta\gamma}$
 is the Levi-Civita symbol, and $\langle i,j\rangle $ in (1)
represents  neighbouring pairs.\\

The Pauli spin martices being representatives of spin-1/2, $\sigma^z$
has got two eigenvalues ($\pm 1$) corrosponding to spins aligned either along 
   z-direction or along the opposite direction respectively.
The eigenstate corrosponding to eigenvalue (+1)  is symolically denoted by
  $|\uparrow\rangle$,
 while that corrosponding to ($-1$) is denoted by   $|\downarrow\rangle$ .\\

If we represent                     
$$ 
 |\uparrow\rangle \Leftrightarrow  \left(\begin{array}{c}
 1\\
0
\end
{array}\right) $$ and 
$$ |\downarrow\rangle \Leftrightarrow \left(
\begin{array}{c}
0\\
1
\end{array}\right), \eqno(4)$$
\noindent then taking these two eigen-vectors as basis, 
Pauli spins have following matrix representations
$$\sigma^x = \left(\begin{array}{cc}
0 & 1\\
1 & 0
\end{array}\right),\hspace{0.2cm}
\sigma^y = \left(\begin{array}{cc}
0 & -i\\
i & 0
\end{array}\right),\hspace{0.2cm} \sigma^z = \left(\begin{array}{cc}
1 & 0\\
0 & -1
\end{array}\right). \eqno (5)$$
 
\noindent With these, one can see that relations in (3) are easily 
satisfied and the tunnelling required in (1) can be easily accommodated.
The order parameter for such a system is generally taken
 to be the expeectation value 
of z-component of the spin, i.e. $\langle\sigma^z\rangle$.
Needless to say that in such a system absolute ordering (complete alinement
 along z-direction ) is not possible even at zero-temperature,  
i.e.,$\qquad \langle\sigma^z\rangle_{T=0} \ne 1 $, when $\Gamma \ne 0$. In
general, therefore, the order $(\langle\sigma^{z}\rangle \ne 0)$ to disorder   
$\langle\sigma^{z}\rangle = 0$ transition can be brought about by tuning 
 either of, or both of the tunnelling field $\Gamma$ and the temperature $T$
(see Fig. 1.2).

 \section{Mean Field Theory (MFT)} 

\leftline{\underline{\textbf{(a) For T = 0}}} 

\vspace{0.35cm}

 Let,
$$\sigma_i^z =  |\vec \sigma|\cos\theta,
\quad \mathrm{and} \quad \sigma_i^x = 
|\vec \sigma|\sin\theta, \eqno (6)$$
\noindent where $\theta$ is the angle between $\vec \sigma$ and z-axis.
 This renders the two mutually non-commuting part of the Hamiltonian (2)
 commuting, since both are expressed in terms of $|\vec \sigma|$ 
operator only. 
 If $\sigma$ is the eigen-value of $|\vec \sigma|$ ($\sigma = 1$ for Pauli 
spin), then the energy per site of the semi-classical system is given by
\cite{BC:DeGenn} 
$$E = -\sigma{}\Gamma\sin\theta -\sigma^2J(0)\cos^2\theta, \eqno (7) $$
\noindent $ J(0) = J_{i}(0) = \sum_{\langle i,j \rangle} J_{ij} $, 
where $j$ indicates the $j$-th nearest 
neighbour of the $i$-th site.
 And the average of the spin-components are given by
\begin{eqnarray*}
\langle\sigma^z \rangle &=& \cos\theta \\
\langle\sigma^x \rangle &=& \sin\theta.
\end{eqnarray*}
\noindent The energy (7) is minimized for $$ \sin{\theta} = 
\Gamma/J(0) \qquad \mathrm{or,} \qquad 
\cos{\theta} = 0. \eqno (8) $$
\noindent Thus we see that if $\Gamma =  0 $, 
$\langle \sigma^x \rangle = 0 $ and the order parameter $\langle \sigma^z 
 \rangle = 1 $, indicating perfect order.

On the other hand, if $ \Gamma < J(0) $, then the ground state is partially 
polarized, since none of $\langle \sigma^z \rangle $ or $ \langle \sigma^x
 \rangle $ is zero.
 However, if $ \Gamma \ge  J(0) $,
 then  we must have 
$ \cos\theta = 0 $ for the ground state energy, which means $\langle
\sigma^z \rangle = 0 $, i.e., the state is a completely disordered
one. Thus, as $ \Gamma $ increases from 0 to J(0), the system undergoes a 
 transition from ordered (ferro)- phase with order parameter
$\langle \sigma^z 
 \rangle = 1 $ to disordered (para)-phase with order parameter 
$\langle \sigma^z \rangle = 0 $ (see Fig. 1.2).

\vspace{0.7cm}
\leftline{\underline{\textbf{ (b) For T $ \ne $ 0 }}}

\vspace{0.35cm}

\begin{figure}
\resizebox{12.0cm}{!}{\rotatebox{270}{\includegraphics{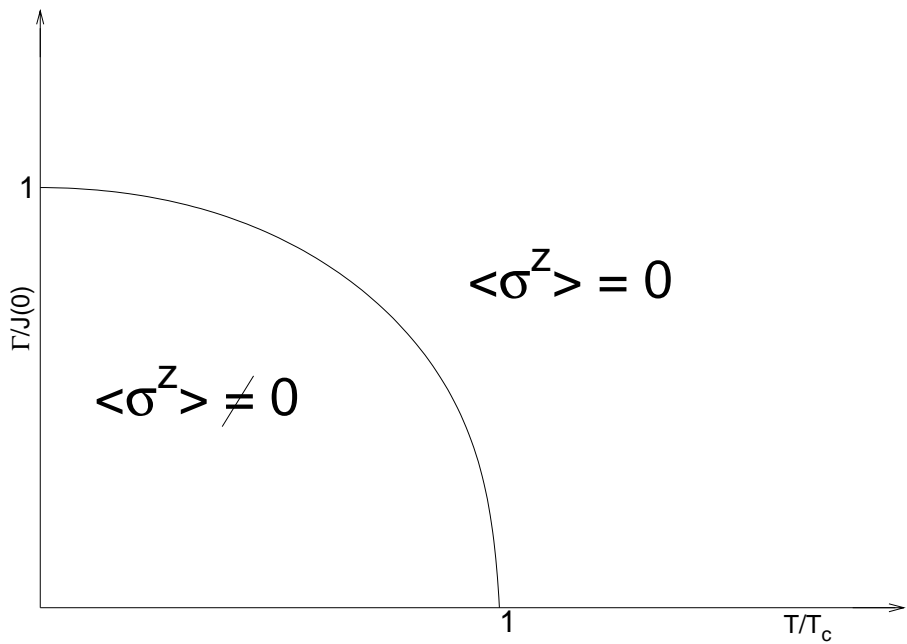}}}
\caption{\small{Schematic phase diagram of the model represented by 
Hamiltonian (2).}}
\end{figure}

\noindent The mean field method can also be extended to\cite{BC:Muller,BC:BKC} 
obtain the behaviour of this
model at non-zero temperature. In this case we define a mean field 
$\mathbf{\vec h}_i$ at each site $i$, which is, in some sense, 
a resultant of the average 
cooperative enforcement in z-direction and the applied transverse field in 
x-direction. Precisely, we take, for general random case, 
  $$ \mathbf{\vec h}_i = \Gamma\hat x + \left(\frac{1}{2}
\sum_j J_{ij}\langle\sigma_j^z\rangle\right)\hat z, \eqno (9) $$
and the spin-vector at the $i$-th site follows $\mathbf{\vec h}_i.$
The spin-vector at $i$-th site is given by
$$ \vec {\mathbf\sigma}_i = \sigma_i^x \hat x + \sigma_i^z \hat z, $$ 
and Hamiltonian thus reads
$$\mathcal{H} = -\sum_i \mathbf{\vec h}_i . \mathbf{\vec \sigma}_i.
 \eqno (10) $$

 For non-random case, all the sites have identical ambience, hence
 $\mathbf{\vec h_i} $ is replaced by  
 $ \mathbf{\vec h} = \Gamma\hat x + \langle \sigma^z
 \rangle J(0) $. And the resulting Hamiltonian takes the form  
$$ \mathcal{H} = -\mathbf{\vec h } . \sum_i\mathbf{\vec 
\sigma}_i. $$  
\noindent The spontaneous magnetization can readily be written down as  
$$ \mathbf{\vec \sigma} = \tanh (\beta|\mathbf{\vec h}|). 
\frac{\mathbf{\vec h}}{|\mathbf{\vec h}|} $$
  $$|\mathbf{\vec h}| = \sqrt{\Gamma^2 + (J(0)\langle \sigma^z \rangle)^2}. 
\eqno (11) $$
Now if $\mathbf{\vec h} $ makes an angle $ \theta $ with z-axis,
 then $\cos\theta =  J(0)\langle \sigma^z\rangle /
  |\mathbf{\vec h}|  $
\quad     and \quad $\sin\theta = \Gamma |\mathbf{\vec h}| $, and hence we 
have 
$$ \langle \sigma^z \rangle = |\mathbf{\vec h}|\cos\theta = 
[ \tanh(\beta|\mathbf{\vec h}|)] \left( \frac{J(0)\langle 
\sigma^z \rangle }{|\mathbf{\vec h}|} \right),  $$
and $$ \langle \sigma^x \rangle = [\tanh (\beta|\mathbf{\vec h}|)]
\frac{\Gamma}{|\mathbf{\vec h}|}. \eqno (12) $$
Here, $\beta =(1/k_B T).$ 
Equation (12) is the self-consistency equation which
can be solved or graphically or otherwise, to obtain the
order parameter $ \langle \sigma^z \rangle $ at any temperature $T$
 and transverse
field $\Gamma $ .
Clearly, the order-disorder transition is tuned both by $\Gamma$ and $T$ 
(see Fig. 1.2).
\vspace{0.7cm}
 
\leftline{\textbf{\underline{$\Gamma = 0$ (Transition driven by $T$): }}}

\vspace{0.35cm}

Here,
$$ \langle \sigma^z \rangle = \tanh \left( \frac{J(0) \langle \sigma^z \rangle
}{k_B T} \right) $$ and $$ \langle \sigma^x \rangle = 0 $$
One can easily see graphically, that the above equations
 has a nontrivial solution
only if $ k_B T < J(0),$ i.e.,\\ 
 
$ \langle \sigma^z \rangle \ne 0  $ \qquad for \quad\ $ k_B T < J(0) $ \\

$ \langle \sigma^z \rangle = 0 $ \qquad for \quad $ k_B T > J(0). $ \\ 
 
 \noindent This shows that there is a critical temperature $ T_c = J(0) $ above which, 
there is no order. 

\vspace{0.8cm}

 \leftline{\textbf{\underline{For $k_B T \rightarrow 0$ 
(Transition driven by $\Gamma$): }}}

\vspace{0.35cm}

Here,
$$\langle \sigma^z \rangle = \frac{J(0) \langle \sigma^z \rangle}
{\sqrt{(\Gamma)^2 + (J(0) \langle \sigma^z \rangle)^2}} \qquad \left(
{\mathrm since,}\quad  \tanh x \Big |_{x \rightarrow \infty } = 1 \right).$$

\noindent From this equation we easily see that in the limit 
$ \Gamma/J(0) \rightarrow 1 $, the only real nontrivial solution is 
$$ \langle \sigma^z \rangle \rightarrow 0 $$  
and $$ \langle \sigma^x \rangle = \frac{\Gamma}{\sqrt{(\Gamma)^2 + (J(0)
\langle \sigma^z \rangle)^2}} \rightarrow 1,\qquad 
\mathrm{as}\quad \frac{\Gamma}{J(0)} \rightarrow 1. $$ 
 
 Thus we see that their is a critical transverse field $ \Gamma_c = 
J(0) $ such that for any $ \Gamma > \Gamma_c $ there is no order even at 
zero temperature.  
In general one sees that at any temperature $ T < T_c $, there exist some 
transverse field $ \Gamma_c $ at which the transition from the ordered
state  $(\langle \sigma^z \rangle \ne 0) $ to the disordered state $(\langle
 \sigma^z \rangle = 0 )$ occurs. The equation for the phase boundary in
the $ (\Gamma-T)$ - plane is obtained by putting
$ \langle \sigma^z \rangle \rightarrow 0 $ in equation (12).
  The equation gives the relation between $\Gamma_c $ and 
$ T_c $ as follows 
$$ \tanh \left(\frac{\Gamma_c}{k_B T} \right) = 
\frac{\Gamma_c}{J(0)}. \eqno (13) $$
One may note that for ordered phase, since $\langle \sigma^z \rangle \ne 0 $,

$$ \frac{1}{| \mathbf{\vec h} |} \tanh (\beta |\mathbf{\vec h}|) = 
\frac{1}{J(0)} = \mathrm{Constant}. $$
\noindent Hence, $ \langle \sigma^x \rangle 
= \left(\Gamma/|\mathbf{\vec h}|\right) \tanh (\beta |\mathbf{\vec h}|) 
= \Gamma/J(0);$
 independent of temperature in the ordered phase. 
While for the disordered phase, since  $ \langle \sigma^z \rangle = 0 $,
$$ \langle \sigma^x \rangle = \tanh (\beta \Gamma ). $$
Using magnetic mapping, mean field theory of this type was indeed
applied to (the BCS theory of) superconductivity \cite{BC:Anderson}, as shown in
appendix A.

\section{Dynamic mode-softening picture}

 The elementary excitations in such a system as described 
above are known as spin waves, and 
they can be studied using Heisenberg equation of motion for $\sigma^z $
using the Hamiltonian. The equation of motion is then given by

$$ \dot \sigma_i^z  = (i\hbar)^{-1}[\sigma_i^z, \mathcal{H} ] \eqno (14) $$ 
or,
$$\dot \sigma_i^z = 2\Gamma \sigma_i^y \qquad (\mathrm{with} \quad \hbar = 1)$$ 

\noindent Hence,
$$ \ddot \sigma_i^z = 2\Gamma \dot \sigma_i^y = 4\Gamma \sum_j J_{ij}\sigma_i^z
\sigma_i^x - 4\Gamma^2  \sigma_i^z.  \eqno (15) $$
With Fourier transforms and random phase approximation 
($ \sigma_i^x \sigma_j^z = \sigma_i^x \langle \sigma_j^z \rangle 
+ \langle \sigma_i^x \rangle \sigma_j^z $ , with $ \langle \sigma^z \rangle = 0 $ in para phase), we get
$$ \omega_q^2 = 4\Gamma (\Gamma - J(q)\langle \sigma^x \rangle), \eqno (16) $$
for the elementary excitations (where $J(q)$ is the 
Fourier transform of $J_{ij}$ ).
 The mode corrosponding to ($ q = 0 $) softens,
i.e., $\omega_0 $ vanishes at the same phase boundary given by 
equation (13).

\section{Suzuki-Trotter Formalism}

Exact analysis for the quantum fluctuation can indeed be tackled by using
renormalization group theory; see appendix B 
for \index{real space!quantum RG}real space quantum RG
theory for one dimensional chain (cf \cite{BC:Jullien}).
 However, such formalisms have
serious limitations in applicability and the 
\index{Suzuki-Trotter!formalism}Suzuki-Trotter formalism 
to map the quantum problem to a classical one has been of enormous practical
importance (e.g. in simulations).

\begin{figure}
 \resizebox{9.50cm}{!}{\rotatebox{270}{\includegraphics{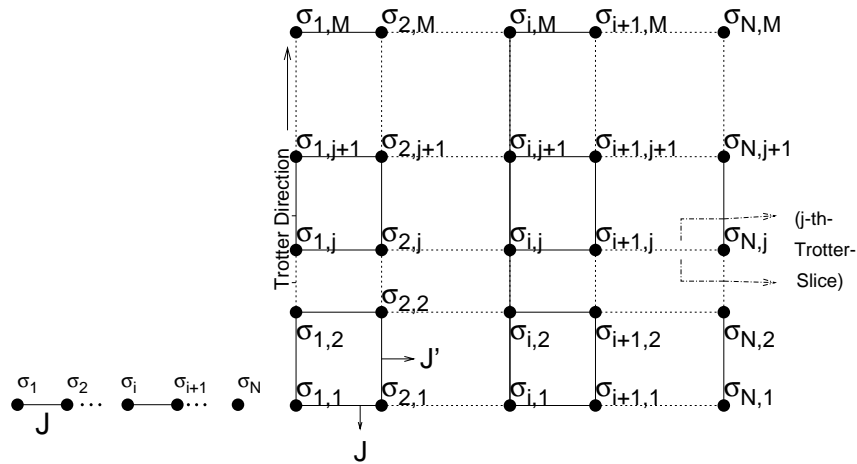}}}
\caption{\small{The Suzuki-Trotter equivalence of quantum one dimensional 
chain and a (1+1) dimensional classical system.
 $J^{\prime}$ indicates the additional
interaction in the Trotter direction.}}
\end{figure}
 \index{Suzuki-Trotter!formalism}Suzuki-Trotter formalism
 \cite{BC:Suzuki} is essentially a method to transform a
$d$-dimensional quantum Hamiltonian into a ($d$+1)-dimensional effective 
classical Hamiltonian giving the same canonical partition function. Let us
illustrate this by applying it to transverse Ising system. 
We start with Transverse Ising Hamiltonian

\begin{eqnarray*}
{\mathcal H} &=& -\Gamma\sum_{i=1}^N \sigma_{i}^{x} - \sum_{(i,j)}J_{ij}
\sigma_{i}^{z}\sigma_{j}^{z} \\ 
 &=& {\mathcal H}_0 + {\mathcal V} \hspace{7cm} (17)
 \end{eqnarray*}

\noindent The canonical partition function of ${\mathcal H}$ reads
$$ Z = Tr e^{-\beta({\mathcal H}_{0}+{\mathcal V})}. $$

\noindent Now we apply the Trotter formula
$$ \exp{(A_1 + A_2)} = \lim_{M\rightarrow \infty} 
\left[ \exp{A_{1}/M}\exp{A_{2}/M} \right]^M, $$
\noindent even when $[A_{1},A_{2}] \ne 0$. On application of this, $Z$ reads
$$ Z = \sum_{i} \lim_{M\rightarrow \infty} 
\langle s_i |\left[ \exp{(-\beta{\mathcal H}_{0}/M)}
\exp{(-\beta{\mathcal V}/M)} \right]^M |s_i \rangle. \eqno (18) $$
\noindent Here $s_i$ represent the $i$-th spin configuration of the whole 
system, and the above summation runs over all such possible configurations 
denoted by $i$. Now we introduce $M$ number of identity operators 
$${\mathcal I} = \sum_{i}^{2^N}|s_{i,k}\rangle\langle s_{i,k}|,\qquad
 k = 1,2,...M.$$
\noindent in between the product of $M$ exponentials in $Z$, and have
$$ Z = \lim_{M\rightarrow\infty} Tr \prod_{k=1}^{M}
\langle \sigma_{1,k}... \sigma_{N,k}|\exp{\left(\frac{-\beta{\mathcal H}_0}{M}
  \right)} \exp{\left(\frac{-\beta{\mathcal V}}{M}\right)}|
\sigma_{1,k+1} ...\sigma_{N,k+1}\rangle, $$ 
\noindent and periodic boundary condition would imply $\sigma_{N+1,p} = 
\sigma_{1,p}$. 
Now, 
$$\prod_{k=1}^{M} \langle \sigma_{1,k}...\sigma_{N,k}|
\exp{\left(\frac{\beta}{M} \sum_{i,j} \sigma_{i}^{z} \sigma_{j}^{z} \right)}|
\sigma_{1,k+1}...\sigma_{N,k+1}\rangle$$
 $$= \exp{\left[\sum_{i,j=1}^{N} \sum_{k=1}^{M}
\frac{\beta J_{ij}}{M} \sigma_{i,k}\sigma_{j,k} \right]}, \eqno (19) $$ 
where $\sigma_{i,k} = \pm 1$ are the eigenvalues of $\sigma^z$ operator. 
Also, 
$$\prod_{k=1}^{M} \langle \sigma_{1,k}...\sigma_{N,k}|
\exp{\left[\frac{\beta \Gamma}{M}\sum_{i} 
\sigma_{i}^{x}\right]}|\sigma_{1,k+1}...\sigma_{N,k+1} \rangle $$ 
$$= \left(\frac{1}{2}\sinh{\left[ \frac{2\beta\Gamma}{M} \right]}
 \right)^{\frac{NM}{2}} \exp{ \left[\frac{1}{2} 
\ln {\coth{\left(\frac{\beta\Gamma}{M}
\right)}}\sum_{i=1}^{N} \sum_{k=1}^{M}\sigma_{i,k}\sigma_{i,k+1} \right]}. 
\eqno (20) $$

\noindent The last step follows because 
 $$e^{a\sigma^x} = e^{-i(ia\sigma^x)} = \cos{(ia\sigma^x)} - i\sin{(ia
\sigma^x)} = \cosh{(a)} + \sigma^{x}\sinh{(a)}, $$
\noindent and therefore
$$\langle\sigma|e^{a\sigma^{x}}|\sigma'\rangle = 
\left[\frac{1}{2}\sinh{(2a)} \right]^{1/2} \exp{\left[(\sigma\sigma'/2) 
\ln{\coth{(a)}}\right]}, $$
\noindent since
$$\langle\uparrow|e^{a\sigma^x}|\uparrow\rangle =  \langle\downarrow
|e^{a\sigma^x}|\downarrow\rangle =  \cosh{(a)} = \left[\frac{1}{2}
\sinh{(2a)}.\coth{(a)} \right]^{1/2}$$
\noindent and
 $$\langle\uparrow|e^{a\sigma^x}|\downarrow\rangle =
\langle \downarrow|e^{a\sigma^x}|\uparrow\rangle =
\sinh{(a)} = \left[\frac{1}{2}\sinh{(2a)}/\coth{(a)}\right]^{1/2}. $$
\noindent Thus the partition function reads 
 $$Z = C^{\frac{NM}{2}}Tr_{\sigma} (-\beta{\mathcal H}_{eff}
[\sigma])\quad ;\quad C = \frac{1}{2}\sinh{\frac{2\beta\Gamma}{M}} $$
\noindent where the effective classical Hamiltonian is
$${\mathcal H}_{eff}(\sigma) = \sum_{(i,j)}^{N} \sum_{k=1}^{M}
\left[-\frac{J_{ij}}{M}\sigma_{i k}\sigma_{j k} - \frac{\delta_{ij}}{2\beta}
\ln{\coth{\left(\frac{\beta\Gamma}{M} \right)}}\sigma_{i k}
\sigma_{i k+1}  \right]. \eqno (21) $$

\begin{figure}
\resizebox{11.0cm}{!}{\rotatebox{270}{\includegraphics{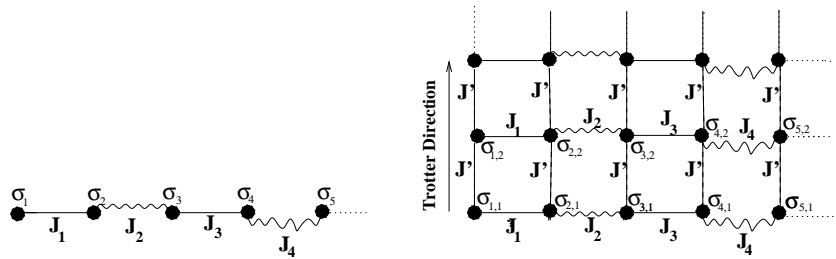}}}
\caption{\small{At the left is a portion of a one dimensional quantum
Ising chain with random exchange interactions and at the right is a part of
its Suzuki-Trotter equivalent classical lattice with randomness correlated in
Trotter direction.}}
\end{figure}
\noindent The Hamiltonian ${\mathcal H}_{eff}$ 
is a classical one, since the variables 
$\sigma_{i,k}$'s involved are merely the eigen-values
 of $\sigma^z$, and hence there is no 
non-commuting part in ${\mathcal H}_{eff}$. 
It may be noted from (21) that $M$ should be at the order of $\hbar \beta$
 (we have taken $\hbar = 1$ in the calculation) for a meaningful comparison
of the interaction in the Trotter direction with that in the original 
Hamiltonian (see Fig. 1.3). For $T\rightarrow 0$, $M\rightarrow\infty$, and  
the Hamiltonian represents a
system of spins in a ($d$+1)-dimensional lattice, which is
one dimension higher than the original $d$-dimensional Hamiltonian,
 as is evident from
the appearence of one extra label $k$ for each spin variable (see Fig. 1.3).
Thus corrosponding to each single quantum spin varible $\sigma_i$
in the original Hamiltonian we have an array of $M$ number of classical
replica spins $\sigma_{i k}$.
This new (time-like)  dimension along which these classical spins are spaced is known as
Trotter dimension.  From the explicit form of ${\mathcal H}_{eff}$, we see that
in addition to the previous interaction ($J$) term 
$(-\sum_{i,j}^{N}J_{ij} \sigma_i
\sigma_j)$, 
there is an additional nearest neighbour interaction ($J'$)
 between the Trotter replicas corrosponding to the same original spin, along
the Trotter direction, given by the term $(\sum_{i,j}^{N}\sum_{k=1}^{M} 
-(\delta_{ij}/2\beta)\ln{\coth{(\beta\Gamma/M)}}
\sigma_{i k}\sigma_{i K+1})$ (as shown in Fig. 1.3). For finite temperature,
the optimal width of the lattice in the Trotter direction is finite
and the critical behaviour remains $d$-dimensional. \\

The calculations, and consequently the effective Hamiltonian (21), is valid for
any general interaction $J_{ij}$; ofcourse, $\Gamma$ has been taken to be 
nonrandom. Fig. 1.4 describes a situation
where $J_{ij}$ were nonrandom (we had $J_{ij} = J$). For random $J_{ij}$,
where $J_{ij}$ were nonrandom (we had $J_{ij} = J$). For random $J_{ij}$,
remain identical ($J^{\prime}$) wheras the spatial randomness in
 interactions for various
Trotter slices get correlated as indicated in Fig. 1.4. Such equivalence
of $d$-dimensional quantum system with a ($d+1$)-dimensional
classical model can also be seen from the renormalization group study of
the quantum models (say, one-dimensional 
\index{transverse Ising model}transverse Ising model
 and its equivalent critical behaviour of 
two-dimensional classical Ising system)
as shown in Appendix B.   
 


 \section{Classical Spin Glasses: A Summary}

 Spin glasses are magnetic systems with randomly competing
(frustrated) interactions \cite{BC:Nishi}. Frustration is a situation where all of the spins
present in the system cannot energetically satisfy every bond associated to 
them. Here the frustration arises due to competing
(ferromagnetic and anti- ferromagnetic) quenched random interactions 
between the spins. As a result there arise huge
barriers ($O(N)$, $N$ = system size) in the free-energy landscape 
of the system. In thermodynamic limit, height of such barriers occassionally 
 go to infinity.
These barrieres strongly separate different configurations
of the system, so that once the system gets stuck in a deep valley in between
two barriers, it practically gets trapped around that configuration for a
macroscopically large time. Because of frustration, the ground state is 
largely degenerate; degeneracy being of the order of $\exp{(N)}$. As discussed
above, these different ground state configurations are often separated by
$O(N)$ barriers, so that once the system settles down in one of them, it 
cannot visit the others equally often in course of time, as predicted by
the Boltzmann probability factor. The system thus becomes ``nonergodic" and
may be described by a nontrivial order parameter distribution \cite{BC:Nishi}
 in the 
thermodynamic limit (unlike the unfrustrated cooperative systems, where the
the distribution becomes trivially delta function-like).  
The spins in such a system thus get frozen in random 
orientations below a certain transition temperature. Although there is no 
long range magnetic order, i.e., the space average of spin moments vanishes,
the time average of any spin is nonzero below the transition (spin-glass) 
temperature. This time average is treated as a measure of spin freezing
or spin glass order parameter. 
    
Several spin glass models have been studied extensively using 
both analytic and computer simulation techniques. The Hamiltonian for such
models can be written as $${\mathcal H} = 
-\sum_{i<j} J_{ij}\sigma_{i}^{z}\sigma_{j}^{z} \eqno (22) $$
\noindent where $S_{i}^{z} = \pm 1,2, ...,N,$ denote the Ising spins, 
interacting with random quenched interactions $J_{ij}$, which differs in 
various models. We will specifically consider
 three extensively studied models. \\

\noindent (a) In Sherrington-Kirkpatrick (S-K) model $J_{ij}$ are long-ranged
 and are distributed with a Gaussian probability (centered around zero), as 
given by
$$P(J_{ij}) = \left(\frac{N}{2\pi J^2} \right)^{1/2}
\exp{\left(\frac{-NJ_{ij}^2}{2J^2}\right)}  \eqno (23) $$

\noindent (b) In Edward-Anderson (EA) model, the $J_{ij}$'s are
short-ranged (say, between the nearest neighbours only), but similarly 
distributed with Gaussian probability (23) \\
 
\noindent (c) In another kind of model, the $J_{ij}$'s are again
short-ranged, but having a binary $(\pm J)$ distribution with probability  $p$:
$$P(J_{ij}) = p\delta (J_{ij} - J) + (1 - p)\delta (J_{ij} - J). \eqno (24) $$ 

 The disorder in the spin system being quenched, one has to 
perform configurational averaging (denoted by overhead bar) over $\ln{Z}$,
where $Z(= Tr \exp{-\beta {\mathcal H}})$ is the partitation function of the system.
To evaluate $\overline{\langle \ln {Z} \rangle}$, one usually employs 
replica trick based on the representation $\ln {Z} = \lim_{n\rightarrow 0}
[(Z^{n} - 1)/n]. $ Now for classical Hamiltonian (with all commuting spin 
components), $Z^{n} = \prod_{\alpha=1}^{n} Z_{\alpha} =
 Z(\sum_{\alpha=1}^{n} H_{\alpha}),$ where $H_{\alpha}$ is the $\alpha$-th
replica of the Hamiltonian ${\mathcal H}$ in equation (22) and $Z_{\alpha}$ is
the corrosponding partition function. The spin freezing can then be measured
in terms of replica overlaps, and Edward-Anderson order parameter takes the 
form
$$ q = \frac{1}{N} 
\sum_{i=1}^{N}\overline{\langle S_{i}^{z}(t)S_{i}^{z}(0)\rangle}|
_{t\rightarrow\infty} \simeq \frac{1}{N}\sum_{i=1}^{N}
\overline{\langle S_{i\alpha }^{z}S_{i\beta }^{z} \rangle}, $$
\noindent where $\alpha $ and $\beta $ corrosponds to different replicas. \\

Extensive Monte Carlo studies, together with the analytical solutions
for the mean field of S-K and EA models, have revealed the nature of spin glass
transition. It appears that the lower critical dimension $d_{l}^{c}$
 for EA model, below which transition ceases to occur (with transition 
temperature $T_c$ becoming zero), is between 2 and 3: $2 < d_{l}^{c} < 3.$
Thu upper critical dimension $d_{u}^{c}$, at and above which mean field
results (e.g., those of S-K model) apply, appears to be 6: $d_{u}^{c} = 6.$
Within these dimensions ($d_{l}^{c} < d < d_{u}^{c}$ ), the spin glass
transitions occur (for Hamiltonians with short-ranged interactions) and
the transition behaviour can be characterized by various exponents. Although 
the linear susceptibility shows a cusp at the transition point, the
nonlinear susceptibility $ \chi_{SG} = (1/N)\sum_{r}g(r)$, where
$g(r) = (1/N)\sum_{i}\overline{(\langle S_{i}^{z}S_{i+r}^{z} \rangle )^2}, $
diverges at the spin glass transition point :
$$\chi_{SG} \sim (T - T_c)^{-\gamma_c},\qquad g(r) \sim r^{-(d-2+\eta_c)}f\left
(\frac{r}{\xi}\right);\qquad \xi \sim |T - T_c|^{-\nu_c}  \eqno (25) $$ 
\noindent Here $\xi$ denotes the correlation length which determines the
length scaling in the spin correlation function $g(r)$ ($f$ in $g(r)$ denotes 
the scaling function). Numerical simulation gives $\nu_c = 1.3\pm 0.1$, 
0.80$\pm$ 0.15, 1/2  and $\gamma_c = 2.9\pm 0.5 $, 1.8$\pm$ 0.4, 1 for
$d = 2,3$ and 6 respectively for the values of exponents. One can define the
characteristic relaxation time $\tau$ through the time dependence of spin 
auto-correlation
$$q(t) = \overline{\langle S_{i}^{z}(t)S_{i}^{z}(0) \rangle } \sim 
t^{-x} \tilde{q}\left( \frac{t}{\tau} \right); \quad \tau \sim \xi^{z}
 \sim |T - T_c|^{-\nu_{c}/z_{c}} \eqno (26) $$
\noindent where $x = (d - 2 + \eta_c)/2z_c$, and $z_c$ denotes the classical 
dynamical exponent. Numerical simulations give $z_c = 6.1\pm 0.3$ and 
$ 4.8\pm 0.4$ in $d =$ 3 and 4 dimensions respectively. Of course, such
large values of $z_c$ (particularly in lower dimensions) also indicates the
 possibility of the failure of power law variation (26) of $\tau$ with
$T - T_c$ and rather suggests a Vogel-Fulcher like variation: $\tau \sim 
\exp{[A/(T - T_c)].}$ In the $\pm J$ spin glasses (type (c) above), some exact
results are known along the `Nishimori Line'\index{Nishimori!line} 
\cite{BC:Nishi}, 
and the nature of the phase
transition there is precisely known.   

\section{Quantum Spin Glasses}

 Quantum spin glasses \cite{BC:Ishi}-\cite{BC:Bhatt} have the 
interesting feature that the 
transition in randomly frustrated (competing) cooperatively interactimg
systems can be driven both by thermal fluctuations or by quantum 
fluctuations. Quantum spin glasses can be of two types: vector spin glasses
introduced by Bray and Moore (see \cite{BC:BKC}), where of course
 quantum fluctuation cannot be
tuned, or a classical spin glass perturbed by some tunable quantum 
fluctuations e.g., as induced by a non commutative transverse
 field \cite{BC:BKC,BC:Ishi}.
The amount of quantum fluctuation being tunable, this Transverse Ising
Spin Glass (TISG) model is perhaps the simplest model in which the quantum 
effects in a random system can be and has been studied extensively and 
systematically \cite{BC:BKC,BC:Bhatt}. Precise realization of TISG in 
${\rm LiHo_{x}Y_{1-x}F_{4}}$, with magnetic Holonium ion concentration around
$x = 0.167$ \cite{BC:Wu}, has led to several important developments. 

The interesting in such quantum spin glass models is about the 
possibility of tunnelling through the (infinitely high) barriers of the
free energy landscape in the classical spin glass models (e.g., S-K model) 
due to the quantum fluctuations induced by the transverse field. In classcal
system, the overriding of an infinitely high barrier is infinitely hard 
for thermal fluctuations at any finite temperature. But quantum fluctuation
can make a system tunnel through such a barrier, if its width is 
infinitessimally small. The barrier widths are actually seen to decrease with
system size indicating to an ergodic (replica symmetric)
picture for the free-energy landscape.   

\subsection{Models}
  
\subsubsection{Sherrington-Kirkpatrick Model in a Transverse Field}
  
\noindent The sherrington-Kirkpatrick (S-K) model in presence of a
 non-commutating tunnelling field, given by the Hamiltonian
$${\mathcal H} = -\sum_{ij}J_{ij}\sigma_{i}^{z}\sigma_{j}^{z} - \Gamma\sum_{i}
\sigma_{i}^{x},
 \eqno (27) $$  
\noindent where the follows the Gaussian distribution
$$P(J_{ij}) = \left( \frac{N}{2\pi\Delta^2} \right)^{1/2}
 \exp{\left( \frac{-NJ_{ij}^{2}}{a\Delta^2}  \right)} \eqno (28) $$
\noindent was first studied by Ishi and Yamamoto \cite{BC:Ishi}.

\vspace{0.55cm}

\leftline{\textbf{Phase Diagram}}

\vspace{0.35cm}

\noindent Several analytical studies have been made to obtain the phase 
diagram of the transverse Ising S-K model (giving in particular the 
zero-temperature critical field). The problem of S-K glass in transverse
field becomes a nontrivial one due to the presence of noncommuting spin
operators in the Hamiltonian. This leads to a dynamical frequency dependent
 (spin) self-interaction.

\vspace{0.35cm}

\leftline{\textit{(i) Mean field estimates :}} 

\vspace{0.35cm}

\noindent One can study an effective spin Hamiltonian for the above 
quantum many body system within the mean field framework. A systematic mean
 field theory for the above model was first carried out
 by Kopec (see e.g., \cite{BC:BKC}), using
 the thermofield dynamical approach and the short time approximation for the 
 dynamical spin self-interaction. Before going into the discussion of this 
approach, we shall briefly review the replica-symmetric solution of the
classical S-K model ($\Gamma = 0$) in a longitudinal field given by the
Hamiltonian $$ {\mathcal H} = -\sum_{\langle ij \rangle}
 J_{ij}\sigma_{i}^{z}\sigma_{j}^{z} - h\sum \sigma_{i}^{z} \eqno (29) $$
\noindent where $J_{ij}$ follows the Gaussian distribution given by (56).
Using the replica trick, one obtains for configuration averaged
 $n$-replicated partition function $\bar{Z}^n$, given by
$$\bar Z^n  = \sum_{(\sigma_{i\alpha} = \pm 1)} \int_{-\infty}^{\infty}
P(J_{ij})dJ_{ij} \exp{\left[ \beta \sum J_{ij}\sum \sigma_{i\alpha}^{z}
\sigma_{j\alpha}^{z} + \beta h\sum \sigma_{i\alpha}^{z}.\right]}$$

\noindent Performing the Gaussian integral, using Hubbard-Stratonovich
 transformation and finally using the method of steepest descent to evaluate
 integrals for thermodynamically large system, one obtains free energy per
site $f$, given by
$$-\beta f = \lim_{n\rightarrow 0} \left[ \frac{\beta\Delta^2}{4} 
\left(1 - \frac{1}{n}\sum_{\alpha , \beta} q_{\alpha ,\beta}^{2} + 
\frac{1}{n}\ln{ Tr (\exp{L})}\right) \right], $$     
\noindent where $L = (\beta J)^2\sum_{\alpha ,\beta} q_{\alpha\beta}
\sigma_{\alpha}^{z}\sigma_{\beta}^{z} 
+ \beta\sum_{\alpha =1}^{n}\sigma_{\alpha}^{z} $ and
 $q_{\alpha\beta}$ is self-consistently given by the saddle point condition
$(\partial f/\partial q_{\alpha\beta }) = 0.$ Cosidering the replica symmetric
 case $(q_{\alpha\beta} = q),$ one finds
$$-\beta f = \frac{(\beta\Delta)^2}{2}(1-q) + 
\frac{1}{\sqrt{2\pi}}\int_{-\infty}^{\infty} dr\quad e^{-\frac{r^2}{2}}
 \ln{[2\cosh{\{\beta h(r)\}}]}$$ 
\noindent where r is the excess static noise arising from the random 
interaction $J_{ij}$ and the spin glass order parameter $q$ is self-consistently given by 
$$ q = \frac{1}{\sqrt{2\pi}}\int_{-\infty}^{\infty} dr\quad  e^{-\frac{r^2}{2}}
\tanh^2{\{ \beta h(r)\}}  $$ 
\noindent and $h(r) = \Delta \sqrt{qr} + h$ can be interpreted as a local
molecular field acting on a site. Different sites have different fields because
of disorder, and the effective distribution of $h(r)$ is Gaussian with mean 0
and varience $\Delta^2 q.$ \\

\noindent At this point we can introduce quantum effect through
 transverse 
field term $-\Gamma\sum_{i}\sigma_{i}^{x}$ 
(with longitudinal field $h = 0$). The
effective single particle Hamiltonian in the 
transverse Ising quantum glass
can be written as
 $${\mathcal H}_s = -h^{z}(r)\sigma^{z} - \Gamma \sigma^{x}, $$
\noindent where $h_{z}(r),$ as mentioned earlier, is the effective field acting
 along the $z$ direction arising due to nonzero value of the the spin glass
order parameter. Treating $h^{z}(r)$ and $\sigma$ as classical vectors in 
pseudo-spin space, one can write the net effective field acting on each spin
as $$h_{0}(r) = h^{z}(r) {\hat z} - \Gamma {\hat x};  \quad |h_{0}(r)| = 
\sqrt{h^{z}(r)^2 + \Gamma^2}. $$
\noindent One can now arrive at the mean field equation for the local 
magnetisation, given by
$$m(r) = p(r) \tanh{[\beta h_{0}(r)]}; \quad p(r) = 
\frac{|h^z (r)|}{|h_{0}(r)|}, $$ 
\noindent and consequently, the spin glass order parameter can be written as
$$ q = \frac{1}{\sqrt{2\pi}}\int_{-\infty}^{\infty} dr\quad e^{-r^2/2}
\tanh^2{\{\beta h_{0}(r) \}}p^2(r). $$ 
\noindent The phase boundary can be found from the above expression by putting
$q \rightarrow 0 \quad (h^{z}(r) = J\sqrt{q}r$ and $h_{0} = \Gamma),$ when
it gives 
$$\frac{\Gamma}{\Delta} = \tanh{\left(\frac{\Gamma}{k_B T}  \right)}. 
\eqno (30) $$
\noindent From above we get $\Gamma_c = J$. Ishi and Yamamoto 
used the ``reaction field"  technique to construct ``TAP" like equation for free
energy of the Hamiltonian (27) and perturbatively expanded the free energy
in powers of $\Gamma$ upto the order $\Gamma^2$ to obtain
$$k_{B}T_{c} = \Delta [1 - 0.23(\Gamma/\Delta)^2]. $$

\vspace{0.35cm}

\leftline{\textit{(ii) \qquad  Monte Carlo Studies :}}

\vspace{0.35cm}
\noindent Several Monte Carlo studies have 
been performed \cite{BC:Ishi},\cite{BC:Bhatt} for S-K spin 
glass in transverse field. Applying \index{Suzuki-Trotter!formalism} 
Suzuki-Trotter formulation
(as discusseed earlier) of effective partition function, 
one can obtain the effective
classical Hamiltonian in $M$th Trotter approximation as
$${\mathcal H}_{eff} =
 -\frac{1}{M}\sum_{i,j=1}^{N}\sum_{k=1}^{M} 
J_{ij}\sigma_{ik}\sigma_{jk} - \frac{1}{2\beta} 
\ln{\coth{\left(\frac{\beta\Gamma}{M}\right)}}
\sum_{i=1}^{N}\sum_{k=1}^{M}\sigma_{ik}\sigma_{i k+1}$$
$$ - \frac{NM}{2}
 \ln{\left[ \frac{1}{2}\sinh{\frac{2\beta\Gamma}{M}} \right]}, \eqno (31)$$
\noindent where $\sigma_{ik}$ denotes the Ising spin defined on the lattice 
($i$, $k$), $i$ being the position in the in the original S-K model and
$k$ denoting the position in the additional Trotter dimension. \\

Ray et al \cite{BC:Ray} took $\Gamma << J$ and their results indeed indicate
a sharp lowering of $T_{C}(\Gamma)$. Such sharp fall of $T_{c}(\Gamma)$ with
large $\Gamma$ is obtained in almost all theoretical studies of the phase 
diagram of the model.

\subsubsection{Edward-Anderson Model in a 
Transverse Field}

The Hamiltonian for the Edward-Anderson spin glass in presence of
transverse field is that given by (27), 
where the random interaction this time
is restricted among the nearest neighbours and satisfies a Gaussian 
distribution with zero mean and variance $J$, as given by
$$P(J_{ij}) = \frac{1}{\sqrt{2\pi}}\exp{\left( -\frac{J^{2}_{ij}}{2J^2}
 \right)}.  $$ 
\noindent With $\Gamma = 0,$ the above model represents the E-A model with
order parameter $q = \overline{\langle \sigma_{i}^{z} \rangle^2} = 1 $ (at $T$ = 0).
When the transverse field is introduced,
 $q$ decreases, and at a critical 
value of the transverse field the order parameter vanishes.
 To study this
quantum phase transition using \index{quantum Monte Carlo} 
quantum Monte Carlo techniques, one must 
remember that the ground state of a $d$-dimensional quantum model is equivalent
to the free enery of a classical model with one added dimension which is the 
imaginary time (Trotter) dimension. The effective classical Hamiltonian can
be written as
$$ {\mathcal H} = \sum_{k}\sum_{ij} K_{ij}\sigma_{ik}\sigma_{jk}
 -\sum_{k}\sum_{i}
K \sigma_{ik}\sigma_{ik+1},  \eqno (32)$$
\noindent with
 $$ K_{ij} = \frac{\beta J_{ij}}{M} ;\qquad K = \frac{1}{2}
\ln{\coth{\left( \frac{\beta \Gamma}{M} \right),}}  $$
\noindent where $\sigma_{ik}$ are classical Ising spins and $(i,j)$ denotes the
original $d$-dimensional lattice sites and $k = 1,2,...,M$ denotes a time
slice. Although the equivalence between classical and the quantum model
holds exactly in the limit $ M\rightarrow \infty , $ one can always make an 
optimum choice for $M$. The equivalent classical Hamiltonian has been
studied using standard Monte Carlo technique. The numerical estimates of the
phase diagram etc. are reviewed in details in \cite{BC:Bhatt}

\subsection{Replica Symmetry in Quantum Spin Glasses}
 The question of existence of replica-symmetric ground states in
quantum spin glasses has been studied extensively in recent years. Replica
 symmetry restoration is a quantum phenomena arising due to the quantum 
tunnelling between the classically `traped' states seperated by infinitely
high (but infinitessimally narrow) barriers in the free energy surface, which
is possible as the tunnelling probability is proportional to the barrier
area, which remains finite. To investigate this aspect of quantum glasses, one has to study the overlap distribution function $P(q)$ given by
$$P(q) = \overline{\sum_{l,l^{\prime}} P_{l}P_{l^{\prime}}
 \delta(q - q^{(ll^{\prime})})}, \eqno (33)  $$ 
\noindent where $P_l$ is the Boltzman weight associated with the state $l$
and $q^{ll^{\prime}}$ is the overlap between the sates $l$ and $l^{\prime}$ 
$$ q^{(ll^{\prime})} = \frac{1}{N}\sum_{i=1}^{N} \langle \sigma_{i}\rangle^{(l)}
\langle \sigma_{i} \rangle^{(l^{\prime})}. \eqno (34) $$
\noindent One can also define the overlap distribution in the following form
(for a finite system of size $N$)
$$P_{N}(q) = \overline{\langle \delta (q - q^{(12)}) \rangle}, \eqno (35) $$
\noindent where $q^{(12)}$ is the overlap between two sets of spins 
$\sigma_{i}^{(1)} $ and $\sigma_{i}^{(2)},$ with identical bond distribution but evolved
with different dynamics,
$$ q^{(12)} = \frac{1}{N}\sum_{i} \sigma_{i}^{(1)}\sigma_{i}^{(2)}. \eqno (36)
 $$
\noindent $P_{N}(q) \rightarrow P(q)$ in the thermodynamic limit. In quantum 
glass problem one can study similarly this overlap distridution $P_{N}(q);$
and if the replica symmetric ground states exists, the above function must tend
 to a delta function in thermodynamic limit. In para-phase, the the
 distribution will approach a delta function
 at $q = 0$ for the infinite system. \\

 Ray, Chakrabarti and Chakrabarti \cite{BC:Ray}, performed Monte Carlo
simulations, mapping the $d$-dimensional transverse
 S-K spin glass Hamiltonian
to an equivalent ($d + 1$)-dimensional classical Hamiltonian and addressed 
the question of stability of the replica symmetric solution, with the choice
of order parameter distribution function given by
$$ P_{N}(q) = \overline{\langle \delta \left( q - \frac{1}{NM}
\sum_{i=1}^{N}\sum_{k=1}^{M}\sigma_{ik}^{(1)}\sigma_{ik}^{(2)} \right)
 \rangle}, \eqno (37) $$    
\noindent where, as mentioned earlier, subscripts (1) and (2) refer to the two 
identical samples but evolved through different Monte Carlo dynamics. It may
be noted that a similar definition for $q$ (involving overlaps in identical
Trotter indices) was used by Guo et al \cite{BC:Guo}. Lai and Goldschmidt 
performed Monte Carlo studies with larger system size ($N \le 100$) and 
studied the order parameter distribution function
$$P_{N}(q) = \overline{\langle \delta \left( q - \frac{1}{N}\sum_{i=1}^{N}
\sigma_{ik}^{(1)}\sigma_{ik^{\prime}}^{(2)} \right) \rangle}, \eqno (38) $$
\noindent where the overlap is taken between different (arbitrarily chosen)
 Trotter indices $k$ and $k^{\prime}$; $k \ne k^{\prime}$. Their studies
indicate that $P_{N}(q)$ does not depend upon the choice of $k$ and
 $k^{\prime}$ (Trotter symmetry). 
Rieger and Young (see \cite{BC:BKC}) also defined $q^{(12)}$ in
similar way $(q^{(12)} = (1/NM)\sum^{i}_{N}\sum_{kk^{\prime}}^{M})
\sigma_{ik}^{(1)}\sigma_{ik^{\prime}}^{(2)}.$
 There are striking differences between the
results Lai and Goldschmidt obtained with the 
results of Ray et al \cite{BC:Ray}. For
$\Gamma << \Gamma_{c}, $ $P(q)$ is found to have (in \cite{BC:Ray})
 an oscillatory dependence on
 $q$ with a frequency linear in $N$ (which is probably due to the formation
of standing waves for identical Trotter overlaps). However, with increase in
$N$, the amplitude of oscillation decreases and the magnitude of $P(q=0)$ 
decreases, indicating that $P(q)$ might go over to a delta function in 
thermodynamic limit. The envelope of this distribution function appears to have
an increasing $P(q=0)$ value as the system size is increased.
 Ray et al \cite{BC:Ray}
argued that the whole spin glass phase is replica symmetric due to quantum
tunnelling between the classical trap states. Lai and Goldschmidt on the other
 hand, do not find any oscillatory behaviour in $P(q)$. In contrary they 
get a \index{replica symmetry!breaking} 
replica symmetry breaking (RSB) in the 
whole spin glass phase from the
nature of $P(q)$, which in this case, has a tail down to $q=0$ even as $N$
increases. According to them their results 
are different from Ray et al \cite{BC:Ray} because
of different choices of the overlap function. Goldschmidt and Lai have also
obtained \index{replica symmetry!breaking} replica symmetry breaking 
solution at first step RSB, and hence the
phase diagram. \\

 B\"{u}ttner and Usadel (see e.g., Chakrabarti et al \cite{BC:BKC}), 
have shown that the replica
 symmetric solution is unstable for the effective classical Hamiltonian (58)
 and also estimated the order parameter and other thermodynamic quantities
like susceptibility, internal energy and entropy by applying Parisi's replica
symmetry breaking scheme to the above effetive Hamiltonian. Using static
approximation, Thirumalai et al (see \cite{BC:BKC}), 
found stable replica symmetric solution
in a small region close to the spin glass freezing temperature near the
phase boundary. But as mentioned earlier, in the region close to 
the critical line, quantum fluctuations are subdued by the thermal
 fluctuations. Thus the restoration of \index{replica symmetry!breaking}
replica symmetry breaking, which is essentially
a quantum effect, perhaps connot be prominent there. \\

 All these numerical studies are for equivalent classical
 Hamiltonian, obtained by applying the \index{Suzuki-Trotter!formalism}
Suzuki-Trotter formalism
 to the original
quantum Hamiltonian, where the interactions are anisotropic in the spatial and
Trotter direction and the interaction in the Trotter direction becomes 
singular in the limit $T\rightarrow 0$. Obviously one cannot extrapolate the
 finite temperature results in zero temperature limit. The results of exact 
diagonalization of finite systems ($N \le 10$) at $T = 0$ itself do not 
indicate any qualitative difference in the behaviour of the 
(configuration average) mass gap $\Delta$ and the internal energy $E_g$ from
that of a ferromagnetic transverse Ising case, indicating the possibility that
 the system might become ``ergodic". On the other hand, the zero temperature
distribution for the order parameter does not appear to go to delta function 
with increasing $N$ as is clearly found for the corrosponding ferromagnet
(random long range interaction without competition). In this case the order
parameter distribution $P(q)$ is simply the number of ground state
 configuratons having the order parameter value as $q$. This perhaps indicate 
broken \index{ergodicity} ergodicity for small values of 
$\Gamma$. The order parameter 
distribution also shows oscillations similar to that
 obtained by Ray et al \cite{BC:Ray}. \\

 Kim and Kim \cite{BC:Kim} have very recently investigated the
 S-K model in transverse field  using imaginary
time replica formalism, under static approximation. They have shown that the
replica-symmetric quantum spin glass phase is stable in most of the area of
the spin glass phase, as have been argued by Ray et al, 
in contrary to the results of Lai et al and Thirumalai et al (see e.g.,
Chakrabarti et al \cite{BC:BKC}).

\section{Quantum Annealing}

\subsection{Multivariable Optimization and Simulated Annealing }
 Multivarable optimization problems consists of finding the 
maximum or minimum values of a function (known as cost function)
 of very many independent variables. A given set of values for the whole
set of independent variables defines a configuration. The value of the 
cost function depends on the configurations, and one has to find the optimum
configuration that minimizes or maximizes the cost function. The explicit
evaluation of the cost function for all possible configurations in this
context, generally turns out to be absolutely impracticable for most systems.

One can therefore start from an arbitrary state and go on changing the 
configuration following some stochastic rule, unless an extremum is reached. 
For example,
in a minimization problem, one may start from an arbitrary configuration, 
change the configuration according to some stochastic rule, evaluate the
cost function of the changed configuration, and then compare its value with
 that of the original configuration. If the new cost function is lower, the
change is retained, i.e.,the new configuration is accepted. Otherwise the
change is not accepted. Such steps may be repeated for times unless a minimum
is reached. But in most cases of multivariable optimisation problem, there are
many local extrema in the cost function landscape, and one cannot be sure that
the extremum that has been reached is the global one. Kirkpatrick et al
\cite{BC:Kirk} proposed a very ingenious physical 
solution to this mathematical problem, 
now known by the name \index{simulated annealing} simulated annealing. 
The basic underlying principle
of \index{simulated annealing} simulated annealing as follows. 
It is known that an ergodic physical
 system, at any finite temperature resides in the global minimum of its free
energy. The minimum of the free energy is a thermodynamic macro-state 
corrosponding to a maximum number of accessible microcsopic configuration.
Hence at thermal equilibrium an ergodic system explores its configuration 
space randomly with equal apriori probability of visiting any configuration,
and consequently is found most of the time at one or other of the 
configurations
that corrosponds to the free energy minimum (since the number of 
configurations corrosponding to such minimum is overwhelmingly large compared
to that of any other macro-state). Now if the system starts from an arbitrary
macrostate (not the minimum of free energy) then due to thermal fluctuation it
reaches the free energy minimum within some time $\tau$ known as the
 thermal relaxation time of the system. 

For an ergodic system
 (away from critical point) this relaxation time increases linearly with
system size (which is logerithmically smaller a number compared to the 
corrosponding number of all possible configurations). Hence if one follows
the random dynamics of the thermal relaxation of a system, then he will be
able to reach the minimum of cost function 
(zero temperatur free energy) in a substancially
smaller time. What one needs to do is to view the cost function $E$ as the
internal energy of some system and start from an
 arbitrary configuration. Then one changes
the configuration according to somestochastic rule, just as before. Now if
the energy is lowered by the change, the change is accepted, but if it is
not, the change is not thrown away with certainity. Instead it is accepted
with a probability equal to the Boltzmann factor $e^{-\Delta E/k_{B}T}$, where
$\Delta E = E_{(\mathrm{after\hspace{0.16cm} change})} - 
E_{(\mathrm{before\hspace{0.16cm} change})}$ (since this is the way how 
systems relax thermally to their free energy minimum). Temperature $T$ here is
an artifically introduced parameter which has a high value initially, and is
reduced slowly as time goes on, finally tending towards zero. At zero
temperature the free energy is nothing but the internal energy of the system,
and thus at the end of the final stage of annealing the
 system can be expected to be
found, with a very high probability, in a configuration that minimizes the
internal energy (cost function). \\

 However this \index{simulated annealing} simulated annealing
 technique can suffer severe set
 back when the system is ``nonergodic", like the spin glasses we discussed
 earlier. In such cases configurations corrosponding to minimum of the
 cost function are separated by $O(N)$ sized barriers, and at any finite 
temperature thermal fluctuations will take practically infinite time to
relax the system to the global minimum crossing these barriers 
in thermodynamic limit $N\rightarrow\infty $.  

\subsection{Ergodicity of Quantum Spin Glasses and Quantum Annealing}
 The \index{non-ergodicity}non-ergodicity problem makes 
the search of the ground state 
of a classical spin glass a computationally hard problem (no algorithm 
bounded by some polynomial in system size exists for such NP-hard problems).
The problems of simulated annealing of 
spin glass-like systems can be overridden (atleast partially) by employing
the method of \index{quantum annealing}quantum annealing 
\cite{BC:Kado,BC:Brook}. 
The basic idea is as follows: First
the problem has to be mapped to a corrosponding physical problem, where the
cost function is represented by some classical Hamiltonian
(say ${\mathcal H}_0$) of the form (22).
Then a suitably chosen noncommuting quantum tunnelling term 
(say ${\mathcal H}^{\prime}(t)$)
 is to be added so that the 
Hamiltonian takes the form of (27). 
One can then solve the time dependent Schrodinger equation
$$i\hbar\frac{\partial \psi}{\partial t} =  
[{\mathcal H}_0 + {\mathcal H}^{\prime}(t)]\psi \eqno (39) $$ for the wave-function 
$\psi(t)$ of the entire system ${\mathcal H}_0 + {\mathcal H}^{\prime}(t)$.
The solution  of the time dependent schrodinger equation approximately 
describes a tunnelling dynamics of the system between different eigenstates
of ${\mathcal H}_0$. 
Like thermal fluctuations in (classical) simulated annealing, 
the quantum (tunnelling) fluctuations owing to ${\mathcal H}^{\prime}$ in (39)
help the system to come out of the local 'trap' states.
If ${\mathcal H}^{\prime}(t) \rightarrow 0$ for 
$t \rightarrow \infty$, the system eventually settles in one of the eigenstates
of ${\mathcal H}_0$; hopefully the ground state.    
The introduction of such a quantum
tunnelling is supposed to make the infinitely high (but infinitessimally thin)
barriers transparent to the system (see, e.g., Appendix C), and it can 
 make transitions to different configurations trapped between such barriers,
 in course of annealing. In other words, it is expected that application of
a quantum tunnelling term will make the free energy landscape ergodic, and 
the system will consequently be able to visit any configuration with finite
probability. Finally the quantum tunnelling term is tuned to zero 
(${\mathcal H}^{\prime}(t) \rightarrow 0$) to get back the classical 
Hamiltonian. It may be 
noted that the success of \index{quantum annealing}quantum annealing
 is directely connected to the
\index{replica symmetry!restoration}replica symmetry restoration
 in quantum spin glass \cite{BC:Ray}, \cite{BC:Kim} due to 
tunnelling through barriers (see Fig. 1.5 and the discussion in the preceeding 
section). \\ 

\begin{figure}
\resizebox{11.0cm}{!}{\rotatebox{270}{\includegraphics{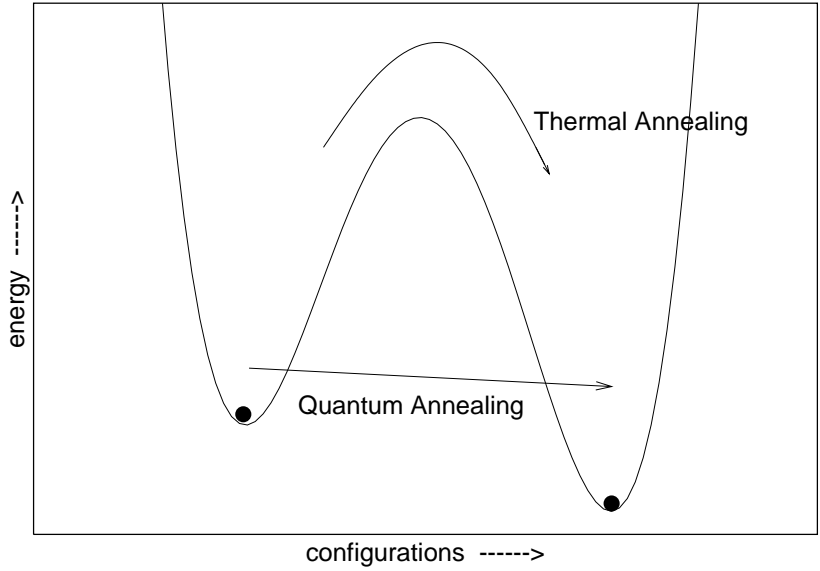}}}
\caption{\small{Schematic indication of the advantage of quantum annealing
over classical annealing.}}
\end{figure}

Here, the $d$-dimensional quantum Hamiltonian (27) 
(to be annealed) is mapped to the $(d+1)$-dimensional effective Hamiltonian

$${\mathcal H}_{d+1} = - \sum_{k+1}^{M}\left( \sum_{i,j}^{N}
J_{ij}\sigma_{i}^{k}\sigma_{j}^{k}
 + J^{\prime}\sum_{i=1}^{N}\sigma_{i}^{k}\sigma_{i}^{k+1} \right), $$
\noindent where $$J^{\prime} =
 -\frac{MT}{2}\ln{\tanh{\left(\frac{\Gamma}{MT}\right)}} > 0 $$ 
\noindent is the nearest neighbour ferromagnetic coupling in Trotter direction,
 between the Trotter replicas of the same spin. 
 In course of annealing, the temperature is kept constant at a low but 
nonzero value, and the tunnelling field $\Gamma$ is tuned slowly 
from a high initial value to zero. The decrease in $\Gamma$ amounts to the
increase in $J^{\prime}$ 
(as casn be seen from above expression of $J^{\prime}$). Initially at high
$\Gamma$, $J^{\prime}$ is low, and each Trotter replica behaves almost
like an independent classical spin system. The tunnelling field is then lowered
in small steps. In each such step, the system is 
annealed in presence of the small temperature. Finally as 
$\Gamma\rightarrow 0$, $J^{\prime}\rightarrow\infty$, forcing all replicas to
coinside at the end.  
 As mentioned already, \index{quantum annealing} quantum annealing
 possibility directly rests on
 the \index{replica symmetry!restoration}replica symmetry restoration
 feature of quantum spin glasses, 
discussed
in earlier section. In fact \index{ergodicity}ergodicity in 
quantum spin glasses, as suggested
in Fig. 1.5 was attributed by Ray et al to the ``quantum fluctuations due 
to transverse field.
Quantum tunnelling between the classical `trap' states, separated by
infinite (but narrow) barriers in the free-energy surface, is possible as
quantum tunnelling probability is proportional to the barrier area which is
finite."  (Sec. V, Ref. \cite{BC:Ray}).

\subsection{Quantum Annealing in Kinetically Constrained Systems}

It is largely believed that apart from the complexity associated to the
non-trivial ground state structure of a glassy system, the occurance of
certain \index{kinetic constraints}kinetic constraints
 (blockings) during relaxation also
contributes substancially
to the slowness of its low temperature dynamics. 
   The Kinetic constraints or blockings can be viewed as infinitely high
 energy barriers appearing in the relaxation path of the system.
 In order to relax to the minimum of the free energy, 
the system has to jump over
 these high barriers thermally, which they fail to do at any finte temperature.
However if such barriers are infinitessimally narrow, then the system might
be able to tunnel through them quantum mechanically if sufficient quantum
fluctuation $\Gamma$ is present in the system. Thus if one tries to anneal 
such a system down to its ground state starting 
from an arbitrary state, then quantum
annealing might turn out to be much superior to the thermal annealing
(see e.g., \cite{BC:Santoro,BC:Bata}).

We have studied \cite{BC:ADas} the annealing of a \index{kinetic constraints}
kinetically constrained
Ising spin chain of $N$ spins, starting from a disordered
state (with negligible initial magnetization),
to its (external field induced) fully ordered ground state.
At any finite temperature $T$ (in the classical model) the system takes an
exponentially long time to relax to the ordered state because of the 
\index{kinetic constraints}kinetic constraints, 
which act like an infinite potential barrier, 
depending on the
neighbouring spin configurations.
Quantum mechanically, this infinite
barrier is taken to be penetrable, ie with finite tunnelling probability, depending
on the barrier height $\chi$ and width $a$ ($a \rightarrow 0$ faster than $\chi^{-2}$).
 The introduced noise, required for the annealing,
 is reduced following an exponential schedule in both the cases:
 $T = T_{0}e^{-t/\tau_C}$, $ \Gamma = \Gamma_{0}e^{-t/\tau_Q},$ with
$T_0 \approx \Gamma_0$. For our simulation for the quantum case,
 we have taken the tunnelling
probabilities $P$ (for cases I-IV) and employed them in a
semi-classical fashion
for the one dimensional spin chain considered.
We observe that for similar achievement
in final order
($m_f \simeq 0.92$ starting from $m_i = 10^{-3}$),
$\tau_C \sim 10^{3}\tau_Q$ for
$N = 5\times 10^{4}$.  For even larger order ($m_f\sim 1$), 
\index{quantum annealing}quantum annealing
works even better ($\tau_C \sim 10^{3}\tau_Q$, for the same value of $N$).
These comparison  are for $g = 10^2$ and $\chi = 10^3$ for the constraint
barriers [20].

In this picture, we considered the collective dynamics of a many particle
system, where each one is confined in a (field) induced asymmetric double
well potential for which we considered only the low lying
 two states (the wave packet localized in
one well or the other), representing the two states (up and down)
of an Ising spin discussed above.
The tunnelling of the wave packet from one well to the other
was taken into account  by employing a scattering picture and we used the
tunnelling probabilities as the flip probabilities
for the quantum Ising spins.
As such, the reported simulation for the one dimensional quantum
East model is a semiclassical one. It may be noted however that,
because of the absence of inter-spin interaction, the dimensionality
actually plays no role in this model except for the fact that the
 \index{kinetic constraints}kinetic constraints
 on any spin depend only on the left nearest
neighbour (directedness in one dimension).
 Hence the semiclassical one dimensional simulation, instead of
a proper \index{quantum Monte Carlo}quantum Monte Carlo simulation
 (equivalent to a higher dimensional classical one
 \cite{BC:BKC}), is quite appropriate here.

\section{Summary and Discussions}

We have introduced the \index{transverse Ising model}
transverse Ising model for discussing the
order-disorder transition (at zero temperature) driven by quantum 
fluctuations. Mean field theories are discussed next in sections 1.3 and 1.4.
 Application to BCS superconductivity theory is discussed in appendix A. 
Renormalization group technique for study of critical behaviour in such quantum 
systems is discussed in appendix B (for a chain). Next we 
have discussed the \index{Suzuki-Trotter!mapping}Suzuki-Trotter mapping of 
the $d$-dimensional quantum
system to $d+1$ dimensional classical system (in section 1.6). We introduce
then the transverse Ising spin glass models, namely, the S-K model in 
transverse field and the
 E-A model in transverse field (Sec. 1.8.2). The existing studies
 on their phase diagrams are discussed briefly. We then discuss about the
problem of \index{replica symmetry!restoration}replica symmetry restoration
in quantum spin glasses 
(in Sec. 1.8.3). The application of the \index{quantum annealing}
quantum annealing
 technique to
capture the near-global minima of NP-hard problems is then discussed, 
and the effectiveness
of quantum tunnelling over the thermal barrier hopping is discussed
(Sec. 1.9).  
  
 It may be noted in this connection that some recent
 attempts have been made to apply similar annealing, induced by quantum 
fluctuations, to the optimization problems like the travelling salesman
problem, image restoration, etc. \cite{BC:Santoro, BC:Bata}. 
Like the near-global minima in 
free energy landscape of such spin glasses, the barriers are often globally 
contributed and  these barrier heights grow as the system size grows (unlike 
the locally optimized configurations and the barriers between them). 
Classically, the system becomes nonergodic due to these macroscopically 
high barriers (NP-hard to reach the ground state), 
as thermal fluctuations have to wait until they can scale such macroscopically
high barriers. Quantum tunnelling does not necessarily look for barrier height
 \cite{BC:Ray} to overcome them 
(by tunnelling through; see appendix C, see Fig. 1.5)
 and helps restoring \index{replica symmetry}replica symmetry 
as well as annealing
\cite{BC:Kado, BC:Brook}.

\bigskip

\leftline{\bf{Acknowledgement:}} 

\noindent We are grateful to Amit Dutta, Jun-ichi Inoue
and Robin Stincombe for many useful discussions and comments.

\section{Appendix} 

\noindent {\Large Appendix A:}

\bigskip

\noindent \index{BCS!Theory of Superconductivity}
{\Large MFT of TIM and BCS Theory of Superconductivity}

\vspace{0.5cm}

 The phonon mediated effective attractive interaction between electrons
give rise to a cooperative quantum Hamiltonian. Although the quantum 
phase transition in such a system is not physical or meaningful, the finite
temperature superconducting phase transition can be studied easily following
the mean field theory discussed here 
(using a pseudo-spin mapping \cite{BC:Anderson}).
 The relevant part of the Hamiltonian of electrons that take part in 
superconductivity has the following form 
$$ \mathcal{H} = \sum_k \epsilon_k (c_k^{\dagger}c_k + c_{-k}^{\dagger}c_{-k})
-V\sum_{k{k^{\prime}}}c_{k^{\prime}}^{\dagger} c_{-k^{\prime}}^{\dagger}c_{-k}
c_{k} \eqno (A1) $$

Here the suffix $k$ indicates a state with momentum $ \vec k $
 and spin up, while ($-k$) indicates a state with momentum
 $-\vec k $ and spin down and V is a positive constant that
models the attractive coupling between electrons through phonons. We will 
solve this equation following spin-analog method \cite{BC:BKC}. 
Here we are considering low-lying states containing
  pair of electrons ($k$, $-k$). 
For a given $k$, there are two possible states that come into consideration:  
either the pair exists, or it does not. Thus we enter into a spin-like two-state
picture as follows.

Let us introduce the number operator   $\hat n_k = c_k^{\dagger}c_k$ .
 This reduces the 
Hamiltonian (A1) to 
$$ \mathcal{H}_{red} = -\sum_k \epsilon_k(1-\hat n_k -\hat n_{-k})
 -V\sum_{kk^{\prime}}c_{k^{\prime}}^{\dagger}c_{-k^{\prime} }^{\dagger}
c_{-k}c_{k}.
  \eqno (A2) $$
Here we have introduced a term $- \sum_k \epsilon_k  $ with the choice 
$ \sum_k \epsilon_k = 0 $ in mind, for all $k$'s (basically these sums are
over the
 states within energy $\pm \omega_D$ about the fermi level, where $\omega_D$
is the Debye energy)  that partictpates in pair formation.
As stated earlier, if $n_k$ denotes the number of
 electrons in $k$-state, then
we are considering only a subspace of states defined by $n_k = n_{-k},$
where either the both of the states in the pair ($k$,$-k$) are occupied, or
both are empyt . Now if we denote by $ |1_k  1_{-k} \rangle $ a ($k$,$-k$)
pair-occupied
state and  by $ |0_k  0_{-k} \rangle $ an unoccupied one,then
$$ (1 - \hat n_k - \hat n_{-k})|1_k  1_{-k} \rangle  =
  (1-1-1)|1_k  1_{-k} \rangle
 = -|1_k  1_{-k} \rangle , $$ and $$ (1 - \hat n_{k} - \hat n_{-k})
 |0_k  0_{-k}
 \rangle = (1-0-0)|0_k  0_{-k} \rangle = |0_k  0_{-k} \rangle $$
   
\noindent Thus we switch over to our good old pseudo-spin picture 
through the following corrospondences
$$ |1_k 1_{-k} \rangle \Leftrightarrow  | \downarrow\rangle_{k},$$ 
$$ |0_k 0_{-k} \rangle \Leftrightarrow  | \uparrow\rangle_{k}, $$  
$$\mathrm{and}\qquad  (1 -n_k -n_{-k})  \Leftrightarrow  
\sigma_k^z. \eqno (A3)$$
Since  
$$ c_k^{\dagger}c_{-k}^{\dagger}|\uparrow\rangle_{k} = |\downarrow\rangle_{k}, 
\quad c_k^{\dagger}c_{-k}^{\dagger}|\downarrow\rangle_{k} = 0 \quad\&\quad 
c_{-k}c_{k}|\downarrow\rangle_{k} = |\uparrow\rangle_{k}, \quad c_{-k}c_{k}|\uparrow\rangle_{k} = 0, $$
\noindent we immediately identify its corrospondence with
raising and lowering operator
$ \sigma^{+}/\sigma^{-}$ :
 \[ \sigma^{-} = \sigma^x - i\sigma^y = 
\left( \begin{array}{cc}
0 & 0 \\
2 & 0
\end{array}\right) \]
\noindent and $$\sigma^{+} = 
\left( \begin{array}{cc}
0 & 2 \\
0 & 0 
\end{array}\right) $$
\noindent and therefore
$$ c_k^{\dagger}c_{-k}^{\dagger} = \frac{1}{2}\sigma_k^{-}, 
\quad c_{-k}c_k = \frac{1}{2}\sigma_k^{\dagger}. \eqno (A4) $$
Hence in terms of these spin operators, Hamiltonian (A2) takes the form 
$$\mathcal{H} = -\sum_k \epsilon_k \sigma_k^z - \frac{1}{4}V\sum_{kk^{\prime}}
\sigma_{k^{\prime}}^{-}\sigma_k^{+}. \eqno (A5) $$
\noindent Since the term $ \sum_{kk^{\prime}}(\sigma_{k^{\prime}}^x \sigma_k^y
  - \sigma_{k^{\prime}}^y \sigma_k^x ) $ vanishes due to symmetric
 summing done 
over $k$ and $k^{\prime}$, the Hamiltonian finally reduces to 
$$  \mathcal{H} = -\sum_k \epsilon_k \sigma_k^z - 
\frac{1}{4}V\sum_{kk^{\prime}}
(\sigma_{k^{\prime}}^x \sigma_k^x + \sigma_{k^{\prime}}^y \sigma_k^y). 
\eqno (A6) $$
 \noindent To obtain the energy spectrum of the pseudo-spin 
\index{BCS!Hamiltonian}BCS Hamiltonian (A6) 
we apply now the mean field theory developed    
in earlier section.

\vspace{0.35cm}

\leftline{\underline{\textbf{Weiss' Mean Field Solution }}}

\vspace{0.35cm}

\noindent Just as we did in case of TIM (see sec 1.3),
 here also we introduce
an average effective field $ \mathbf{\vec h}_k $ 
for each pseudo-spin $\sigma_{k}$ as 
$$ \mathbf{\vec h}_k = \epsilon_k \hat z + \frac{1}{2}V\sum_{k^{\prime}} 
(\langle \sigma_{k^{\prime}}^x \rangle \hat x
+\langle \sigma_{k^{\prime}}^y \rangle \hat y )  $$ 
and conseqently the Hamiltonian (A6) takes the form
$$ \mathcal{H} = - \sum_k \mathbf{\vec h}_k . \mathbf{\vec \sigma}_k.$$
Here for each $k$ there is an independent spin $\mathbf{\vec \sigma}_k$
which interacts only with some effective 
field $ \mathbf{\vec h}_k $, and our system is a 
collection of such mutually non-interacting spins 
under a field ${\mathbf{\vec h}_k}$. \\

Now if redefine our x-axis along the projection of $ \mathbf{\vec h_k} $ on the
x-y plane for each k, then with all 
$  \langle \sigma_{k^{\prime}}^y \rangle = 0 $ we get
$$ \tan \theta_k = \frac{h_k^x}{h_k^z} = \frac{\frac{1}{2} V
 \sum_{k^{\prime}} \langle \sigma_{k^{\prime}}^x \rangle}
{\epsilon_k} \quad , \eqno (A7) $$  
where $ \theta_k $ is the angle between z-axis and $ \mathbf{\vec h}_k  $.

\vspace{0.35cm}

\leftline{\underline{\textbf{Excitation spectra at T = 0}}}

\vspace{0.35cm}

\noindent Since at $T = 0$ $\langle\sigma^{x}\rangle = 1$,
$$\langle \sigma_{k^{\prime}}^{x} \rangle = |\mathbf{\vec \sigma}| 
\sin \theta_{k^{\prime}} =  \sin \theta_{k^{\prime}} \eqno (A8) $$
Thus from (A7) we get
$$ \tan \theta_k = (v/2\epsilon_k) \sum_{k^{\prime}} \sin \theta_{k^{\prime}}
 $$
Now let us define $$ \Delta \equiv \frac{1}{2} V \sum_{k^{\prime}} 
\sin \theta_{k^{\prime}}, $$ 
so that $ \tan \theta_k = \Delta/\epsilon_k $.
Then simple trigonometry gives -
$$ \sin \theta_k = \frac{\Delta}{\sqrt{\Delta^2 + \epsilon_k^2}}\quad;
\qquad \cos \theta_k = \frac{\epsilon_k}{\sqrt{\Delta^2 + \epsilon_k^2}}.
\eqno (A9) $$
Substituting for $ \sin \theta_{k^{\prime}} $ into the above equation we get
$$ \Delta = \frac{1}{2}V \sum_{k^{\prime}} \frac{\Delta}{\sqrt{\Delta^2 + 
\epsilon_{k^{\prime}}^2}}. $$
Assuming the spectrum to be nearly continuous, 
we replace the summation by an
integral and note that V is attractive for energy 
within  $ \pm \omega_D $ on both
sides of fermi level; $ \omega_D $ being of the order of Debye energy. then 
the last equation becomes
$$ 1 = \frac{1}{2}V\rho_F \int_{-\omega_D}^{\omega_D} \frac{d\epsilon}{
 \sqrt{\Delta^2 + \epsilon^2}} = V\rho_F \sinh^{-1} (\omega_D/\Delta).$$  
Here $\rho_F  $ is the density of states at fermi level.
Thus $$ \Delta = \frac{\omega_D}{\sinh (1/V \rho_F )} 
 \cong 2\omega_D e^{-1/V\rho_F},\qquad(\quad \mathrm{if} 
\quad \rho_F V << 1 ) 
\eqno (A10) $$
We see that $\Delta $ is positive if V is positive. To interprate the parameter
$ \Delta $, one may notice that at first approximation, the excitation spectrum
 is obtained as the energy $\mathcal{E}_k $ to reverse a pseudo- spin in the
field $ \mathbf{\vec h}_k $ , i.e.,
$$ \mathcal{E}_k = 2|\mathbf{\vec h}_k| = 2\left( \epsilon_k^2 + \Delta^2
 \right)^{1/2}. \eqno (A11) $$
From this expression we clearly see that the minimum excitation energy is
$ 2\Delta $, i.e. $ \Delta $ gives the energy gap in the excitation spectrum.\\

\vspace{0.35cm}

\noindent {\underline{\bf Estimating transition temperature $T_c$}}

\vspace{0.35cm}

\noindent To find the critical temperature for
 \index{BCS!transition}BCS transition,
 we just extend here the 
non-zero temperature version of mean field theory done for Ising case. 
We should have (unlike that in (A11), where $\langle\sigma_{k}\rangle = 1$)
for $T = 0$: 
 
$$  \langle \mathbf{ \sigma}_k^z \rangle = 
\tanh \left(\beta |\mathbf{\vec h}_k|\right). \eqno (A12) $$
Equation (23) accordingly modifies to 
$$ \tan \theta_k = \left( \frac{V}{2\epsilon_k} \right)
 \sum_{k^{\prime}} \tanh \big(\beta
 |\mathbf{\vec h}_{k^{\prime}}| \big) \sin \theta_{k^{\prime}} 
\equiv \frac{\Delta(T)}{\epsilon_k},  \eqno (A13)  $$
\noindent where $\Delta(T) = \frac{V}{2}\sum_{k^{\prime}}
 \tanh{\left(\frac{|{\vec h}_{k^{\prime}}|}{T}
 \right)}\sin{\theta_{k^{\prime}}}. $ 
From equation (A11) we have 
$$ |\mathbf{\vec h}_k| = \big[ \epsilon_k^2 + \Delta^2(T) \big].
 $$
The \index{BCS!transition}BCS transition
 is characterized by the vanishing of the 
gap $ \Delta$,
since without such a gap in the spectrum, infinite conductance would not be
 possible except at $T = 0$. 
Hence, as $T \rightarrow T_c,\quad \Delta \rightarrow 0 $, 
i.e., using (A11), 
$ |\mathbf{\vec h}_k| = \epsilon_k $ and putting this and relations like (A9) 
in (A13), we get
$$ 1 = V\sum_{k^{\prime}} \frac{1}{2\epsilon_{k^{\prime}}} 
\tanh \left( \frac{\epsilon_{k^{\prime}}}{T_c} \right). \eqno (A14) $$
Above relation is correct if we consider an excited pair as a single entity.
However, if we extend our picture to incorporate single particles excited 
symmetrically in momentum space, then we double the number of possible
 excitations, thereby doubling the overall entropy. This is exactly equivalent
to a doubling of the temperature in free energy. The energy contribution to the
free energy, however, remains unaltered, since two single particle excitations
of same $ \mathbf{|\vec k|}  $ have same energy as that of a pair of equal 
$ \mathbf{|\vec k|}$. Hence we replace $ T_c $ by $ 2T_c $, and in the continuum
limit, get
 
$$ \frac{2}{V\rho_F} = \int_{-\omega_D}^{\omega_D} \frac{d\epsilon}{\epsilon}
\tanh \left(\frac{\epsilon}{2T_c}\right) = 2\int_0^{\omega_D/2T_c} 
 \frac{\tanh x}{x} dx,$$ 
with $ (x =\epsilon/2T_c )$
This is the equation from which we obtain $ T_c $ on integration.
If $ T_c << \omega_D $, then we may approximate $\tanh x 
\approx 1$, for $ x \ge  1 $, and for $x << 1$,
 we set $\tanh x \approx
x $. This readily reduces the integral to the value 
$ 1 + \log (\omega_D/2T_c),
 $ from  which we have $$ T_c = (e/2)\omega_D e^{-1/V\rho_F}.$$
 Grphical integration gives a closer
result
$$ T_c = 1.14\omega_D e^{-1/V\rho_F}. \eqno (A15) $$
Comparing  equations (A10) and (A15) we get the approximate relationship 
$$ 2\Delta \simeq 3.5 T_c.  \eqno (A16) $$ 
This result is quite consistent with the exprimental values for a number of 
materials. For example, the value of $ 2\Delta/T_c $ are 3.5, 3.4, 4.1, 
 3.3 for Sn, Al, Pb, and Cd superconductors respectively.

\vspace{0.5cm}

\noindent {\Large Appendix B:}

\bigskip

\noindent \index{real space!renormalization}
{\Large Real Space Renormalization for Transverse Ising Chain}

\vspace{0.5cm}

 Here the basic idea of \index{real space!block renormalization}real space
block renormalization 
\cite{BC:Jullien}, \cite{BC:BKC} is
illustrated by applying it on an Ising chain in transverse field. Taking the
cooperative interaction along x-axis, and the transverse field along z-axis,
the Hamiltonian reads
\begin{eqnarray*}
{\mathcal H} &=& -\Gamma\sum_{i=1}^{N}\sigma_{i}^{z} 
-J\sum_{i=1}^{N-1}\sigma_{i}^{x}\sigma_{i+1}^{x}  \\
 &=& {\mathcal H}_B + {\mathcal H}_{IB} \quad {\mathrm {(say)}. \hspace{7.34cm}
 (B1)}
\end{eqnarray*}
\noindent Here $${\mathcal{H}}_B = \sum_{p=1}^{N/b} {\mathcal H}_p 
\quad ; \qquad {\mathcal H}_p = -\sum_{i=1}^{b} \Gamma \sigma_{i,p}^{z} - 
\sum_{i-1}^{b-1} J \sigma_{i,p}^{x}\sigma_{i+1,p}^{x} \eqno (B2) $$
\noindent and
$$ {\mathcal H}_{IB} = \sum_{p=1}^{N/(b-1)} {\mathcal H}_{p,p+1}\quad; \qquad
{\mathcal H }_{p,p+1} = -J\sigma_{b,p}^{x}\sigma_{1,p+1}^{x}. \eqno (B3) $$ 
\noindent The above rearrangement of the Hamiltonian recasts the 
picture of $N$ spins with nearest-neighbour interaction into one in which there
are $N/(b-1)$ blocks, each consisting of $b$ number of spins. The part 
${\mathcal H}_{B}$  represents the interaction between the spins within 
the blocks, while ${\mathcal H}_{IB}$ represents the interactions between the
blocks through their terminal spins(see Fig. 1.6). \\

\begin{figure}
\resizebox{12.0cm}{!}{\rotatebox{270}{\includegraphics{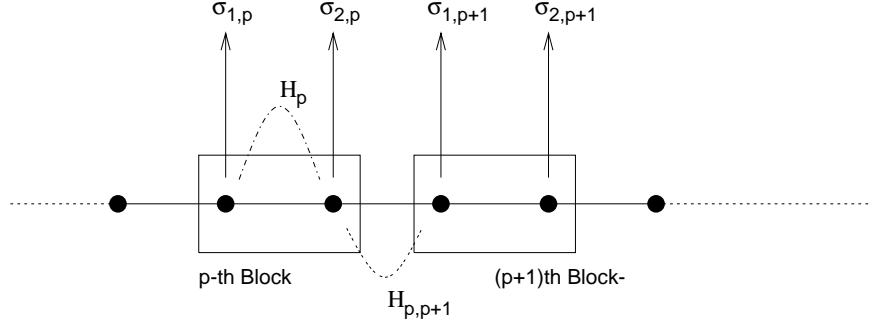}}}
\caption{\small{The linear chain is broken up into blocks of size
 $b$ ($=2$ here)
 and the Hamiltonian (B1) can be written as the sum of block Hamiltonians 
${\mathcal H}_p$ and inter-block Hamiltonians ${\mathcal H}_{p,p+1}$. The
Hamiltonian ${\mathcal H}_p$ is diagonalized exactly and the lowest lying two
states are identified as the renormalized spin states in terms of which the
inter-block Hamiltonian is rewritten to get the RG recursion relation. }}
\end{figure}
\noindent Here we will consider $b=2$, as shown in the figure. Now 
${\mathcal H}_{p}$ has got 4 eigen-states, and one can express them in terms
of the linear superposition of the eigen-states of
 $\sigma_{1,p}^{z}\otimes \sigma_{2,p}^{z}$; namely,
$$|\uparrow \uparrow\rangle ,\quad |\downarrow \downarrow\rangle ,
\quad |\uparrow \downarrow\rangle ,\quad \mathrm{and}\quad 
|\downarrow \uparrow\rangle . $$  
\noindent Considering the orthonormality of the eigen-states, one may easily
see that the eigenstates of ${\mathcal H}_{p}$ can be
expressed as
\begin{eqnarray*}
|0\rangle &=& \frac{1}{\sqrt{1+a^2}}(|\uparrow\uparrow\rangle + 
a|\downarrow\downarrow\rangle) \\
|1\rangle &=& \frac{1}{\sqrt{2}}(|\uparrow\downarrow\rangle 
+ |\downarrow\uparrow\rangle) \\
|2\rangle &=& \frac{1}{\sqrt{2}}(|\uparrow\downarrow\rangle - 
|\downarrow\uparrow\rangle) \\
|3\rangle &=& \frac{1}{\sqrt{1+a^2}}(a|\uparrow\uparrow\rangle 
- |\downarrow\downarrow\rangle). \hspace{6.5cm} (B4)
\end{eqnarray*} 
\noindent Here $a$ is a coefficient required to be chosen properly, so that
$|0\rangle$ and $|3\rangle$ are eigenstates of $\mathcal{H}_p$. One gets,
\begin{eqnarray*}
{\mathcal H}_P|0\rangle &=& {\mathcal H}_p 
\left[\frac{1}{\sqrt{1 + a^2}}|\uparrow\uparrow\rangle + a|\downarrow\downarrow
\rangle\right] \\
 &=& [-\Gamma(\sigma_{1}^{z} + \sigma_{2}^{z})  
- J(\sigma_{1}^{x}\sigma_{2}^{x})]\frac{1}{\sqrt{1 + a^2}}
(|\uparrow\uparrow\rangle +a|\downarrow\downarrow\rangle) \\
&=& \frac{1}{\sqrt{1 + a^2}}[-\Gamma(2|\uparrow\uparrow\rangle 
- 2a|\downarrow\downarrow\rangle) -J(|\downarrow\downarrow\rangle) 
+ a|\uparrow\uparrow\rangle] \\
&=& -(2\Gamma + Ja)\frac{1}{\sqrt{1 + a^2}}\left[|\uparrow\uparrow\rangle 
 + \left(-\frac{2\Gamma - J/a}{2\Gamma + Ja} 
\right)a|\downarrow\downarrow\rangle\right]
\end{eqnarray*} 
\noindent Thus $|0\rangle$ to be an eigenstate of ${\mathcal H}_p$, one must
have
$$-\frac{2\Gamma - J/a}{2\Gamma + Ja} = 1 $$
$$=> \qquad Ja^2 - 4\Gamma a - J = 0 $$
$$\mathrm{or},\qquad  a = \frac{\pm \sqrt{4\Gamma^2 + J^2} - 2\Gamma}{J}. 
\eqno (B5)$$
\noindent To minimize the energy, we have to choose,
$$ a = \frac{\sqrt{4\Gamma^2 + J^2} - 2\Gamma}{J}. $$

\noindent One can now see, applying ${\mathcal H}_p$ on its eigen-states,
\begin{eqnarray*}
{\mathcal H}_p |0\rangle &=& E_0 |0\rangle, \quad E_0 = 
-\sqrt{4\Gamma^2 + J^2} \\
{\mathcal H}_p |1\rangle &=& E_1 |1\rangle, \quad E_1 = -J \\
{\mathcal H}_p |2\rangle &=& E_2 |2\rangle, \quad E_2 = +J \\
{\mathcal H}_p |3\rangle &=& E_3 |3\rangle, \quad E_3 =
+\sqrt{4\Gamma^2 + J^2}. \hspace{5.0cm} (B6)
\end{eqnarray*} 

\noindent Now we define our new renormalized spin variables 
$\sigma^{\prime}$'s, 
each replacing a block in the original Hamiltonian. We retain only the two
lowest lying states $|0\rangle$ and $|1\rangle$ of a block and define 
corrosponding $\sigma_{p}^{\prime Z}$ to have them as its two eigenstates,
 $|\uparrow\rangle = |0\rangle$ and $|\downarrow\rangle = |1\rangle$. We 
also define
$$\sigma^{\prime x} = \frac{\sigma_{1}^{x}\otimes {\mathcal I} + 
{\mathcal I}\otimes \sigma_{2}^{x}}{2}, $$
\noindent where ${\mathcal I}$ is the $2\times 2$ identity matrix. Now since
$$\langle 0|\sigma^{\prime x}|1\rangle = \frac{1 + a}{\sqrt{2(1 + a^2)}}, $$
we take our renormalized $J$ to be
$$ J^{\prime} = J\frac{(1 + a)^2}{2(1 + a^2)}, \eqno (B7) $$
\noindent and since the energy gap between $|0\rangle $ and $|1\rangle $
must be equal to $2\Gamma^{\prime}$ (This gap was $2\Gamma$ in the 
unrenormalized states), we set 
$$ \Gamma^{\prime} = \frac{E_1 - E_0}{2} = 
\frac{\sqrt{4\Gamma^2 + J^2} + J}{2} = 
 \frac{J}{2}[\sqrt{4\lambda^{2} + 1} + 1], \eqno (B8) $$
\noindent where $ a = \sqrt{4\lambda^2 + 1} - 2\lambda.$, defining the 
relevant variable $\lambda = \Gamma/J$. \\

\noindent The fixed points of the recurrence relation (rewritten in terms
 of $\lambda$) are
\begin{eqnarray*}
\lambda^{\star} &=& 0 \\
\lambda^{\star} &\rightarrow& \infty \\
\mathrm{and} \quad \lambda^{\star} &\simeq& 1.277. \hspace{7.7cm} (B9)
 \end{eqnarray*}
   
\noindent Now if correlation length goes as 
$$\xi \sim (\lambda - \lambda_{c})^{\nu}, $$
\noindent in the original system, then in the renormalized system we should 
have 
$$\xi^{\prime} \sim (\lambda^{\prime} - \lambda_{c})^{\nu} $$
$$=> \qquad \frac{\xi^{\prime}}{\xi} = 
\left(\frac{\lambda^{\prime} - \lambda_{c}}{\lambda - \lambda_{c}} \right)
^{-\nu} => \left(\frac{\xi^{\prime}}{\xi}\right)^{-1/\nu} =
 \frac{d\lambda^{\prime}}{d\lambda}
\Big |_{\lambda=\lambda_{c}\equiv \lambda^{\star}}. \eqno (B10) $$
Now since the actual physical correlation length should remain same as we
renormalize, $\xi^{\prime}$ (correlation length in the 
renormalized length scale)
must be smaller by the factor $b$ (that scales the length), than  
 $\xi$ (correlation length in original scale). i.e.,
$\xi^{\prime}/\xi = b,$ or,
$$ b^{-1/\nu} =\left( \frac{d\lambda^{\prime}}{d\lambda}\right)_
{\lambda =\lambda_{c}=\lambda^{\star} } \equiv  \Omega\qquad \mathrm{(say),}$$  
$$\mathrm{hence} ,\qquad \qquad \nu = 
\left(\frac{\ln \Omega}{\ln b}\right)_{\lambda = \lambda^{\star}}
 = \frac{\ln \Omega}{\ln 2} \simeq 1.47, \qquad \mathrm{(for}\quad b = 2), 
\eqno (B11) $$
\noindent compared to the exact value $\nu = 1$ for ($d+1 =)$ 2 dimensional
classical Ising system.
\noindent Similarly $E_g \sim \omega \sim (\mathrm{time})^{-1} \sim \xi^{-z};
 \quad z = 1.$ 
But for $b = 2$, we donot get $z = 1$. 
Instead,  
 $\lambda^{\prime}/\lambda \sim b^{-z} $ gives $z \simeq 0.55$. Energy gap 

$$ \Delta (\lambda) \sim |\lambda_{c} - \lambda|^{s} \sim \xi^{-z}
 \sim |\lambda^{c} - \lambda|^{\nu z} \eqno (B12)  $$ 
\noindent Hence $s = \nu z =  0.55 \times 1.47 \simeq 0.81 $ 
(compared to the exact
result $s$ = 1). Results improve rapidly for large $b$ values \cite{BC:Jullien}.

\vspace{0.5cm}

\noindent {\Large Appendix C:}

\bigskip

\noindent {\Large Tunnelling Through \index{asymmetric barrier}Asymmetric 
Barrier}

\vspace{0.5cm}

Let us consider an asymmetric potential energy barrier in one dimension, 
(as shown in Fig. 1.7). It is essentially a rectangular barrier 
of height $\chi$ and width $a$ between
two different energy levels with a potential difference $h$ between them. 
\begin{figure*}
\resizebox{7.5cm}{!}{\includegraphics{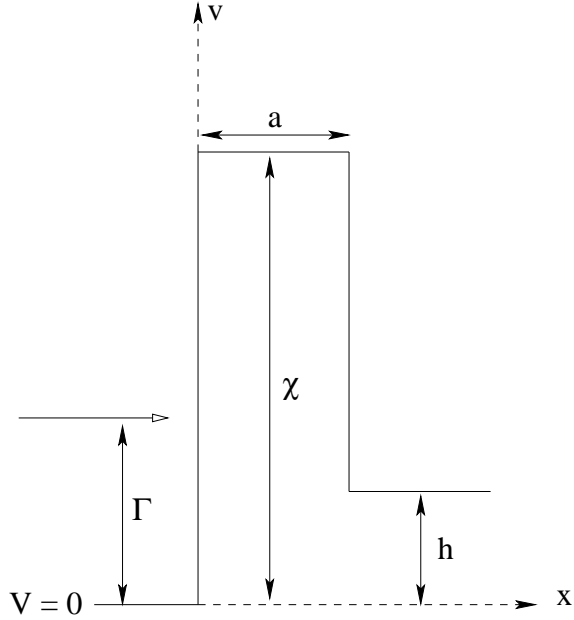}}
\caption{\small{Quantum tunnelling thruogh asymmetric barrier.}}
\end{figure*}
The potential energy $V$ is zero at the left of the barrier ($x < 0$), 
and it is $h$ ($h$ may be negative as well) at the right of the 
barrier ($x > a$). 
If a beam of free particles of mass $m$ with kinetic energy $\Gamma$ 
incidents on the barrier from the left, then one can calculate the
probability for a particle in the beam to get transmitted through
(or reflected by) the barrier by solving the time-independent Schr\"{o}dinger
equation (with a time-independent $V$). The transmission coefficient $T$
(defined below) describes the probability of transmission for a single
particle, as well as the average transmission of the incident beam. 
 
The incident wave function $\psi_{1}(x),$ the intermediate
wave function $\psi_{2}(x)$ and the transmitted
wave function $\psi_{3}(x)$ then takes the form
\begin{eqnarray*}
\psi_{1}(x) &=& A e^{-ik_{1}x}, \qquad x<0, \\
\psi_{2}(x) &=& B e^{k_{2}x} + C e^{-k_{2}x}, \qquad 0\le x\le a\\
\psi_{3}(x) &=& D e^{ik_{3}x}, \qquad x>a
\end{eqnarray*}
\noindent where, 
 $$k_{1}^2 = \Gamma ;\qquad 
   k_{2}^2 = \Gamma - \chi \quad
{\rm and}\qquad k_{3}^2 = \Gamma - h, $$ 
\noindent setting $2m/\hbar^2 = 1$. Here $A$ and $D$ are the 
amplitudes of the incident and the transmitted wave respectively.  
At this point 
one may note that for $\Gamma<h$ transmission is trivially zero. 
Hence we consider the case
for $\Gamma>h$ i.e., for real $k_{3}$. In that case, applying the
condition of continuiety of the wave function and its space derivatives at
the boundaries, one obtains the relation (cf. \cite{BC:Murphy})
$$A = \frac{1}{2}D e^{ik_{3}a}[(1 + k_{3}/k_{1})\cosh{\kappa a} + 
 i(\kappa/k_{1} - k_{3}/\kappa)\sinh{\kappa a}], $$ 
\noindent where $\kappa^2 = -k_{2}^2 = \chi - \Gamma$. We now
consider the limit of very high but narrow barrier, such that
$\chi \rightarrow \infty$, $a \rightarrow 0$, 
with $g = \chi a$ finite. We also assume that
$\chi \gg \Gamma$, so that $\kappa^{2} \approx \chi$, and of course 
$\kappa$ is real. Since $\Gamma \ge 0$, $k_{1}$ is also real. Hence under 
this condition the transmission coefficient defined as 
$T = |D|^{2}k_{3}/|A|^{2}k_{1}$ is given by (cf. \cite{BC:Murphy})
$$T = \frac{4k_{3}/k_{1}}{\left(1 + 
\frac{k_3}{k_1}\right)^{2}\cosh^{2}{(\kappa a)} + 
\left(\frac{\kappa}{k_1} - 
\frac{k_3}{\kappa}\right)^{2}\sinh^{2}{(\kappa a)}}.$$
\noindent In the limit of high but narrow barrier specified above, one
has $\kappa a \ll 1$. Hence neglecting terms quadratic or of higher order
in $\kappa a$ and linear in $1/\kappa$, one gets
\begin{eqnarray*} 
 T &\approx& \frac{4k_{3}/k_{1}}{\left(1 
+ \frac{k_3}{k_1}\right)^{2} +
\left(\frac{\kappa}{k_{1}}\right)^{2}(\kappa a)^2} \\
&=& \frac{4\sqrt{\Gamma(\Gamma - h)}}{(\sqrt{\Gamma} + 
\sqrt{\Gamma - h})^2 + g^2},
\end{eqnarray*} 
\noindent putting $k_{1} = \sqrt{\Gamma},$ $k_{3} = \sqrt{\Gamma - h}$ and
$\kappa^{2}a \approx \chi a = g$. The transmission coefficient $T$ is thus
finite even when the barrier height $\chi$ diverges keeping $g = \chi a$
finite.

%
 
%
%

\nobreak

\end{document}